\documentclass[preprint,showpacs,showemail,preprintnumbers,prb]{revtex4}
\usepackage{graphicx}
\usepackage{dcolumn}
\usepackage{bm}
\usepackage{epsf}

\begin{document}
\title{
Nanostructure and velocity of field-driven solid-on-solid interfaces
moving under a phonon-assisted dynamic
}

\author{
G.~M.~Buend{\'{\i}}a$^{1}$}\email{buendia@usb.ve}
\author{
P.~A.~ Rikvold$^{2,3,4}$}\email{rikvold@scs.fsu.edu}
\author{M.~Kolesik$^{5,6}$}\email{kolesik@acms.arizona.edu}
\author{K.~Park$^{7}$}\email{kyungwha@pontiac.phys.vt.edu}
\author{M.~A.~Novotny$^{8}$}\email{man40@ra.msstate.edu}

\affiliation{
$^1$Department of Physics, Universidad Sim{\'o}n Bol{\'{\i}}var, Caracas 1080, Venezuela\\
$^2$School of Computational Science, Florida State University, Tallahassee, Florida
32306-4120, USA\\
$^3$Center for Materials Research and Technology
and Department of Physics,
Florida State University, Tallahassee, Florida 32306-4350, USA\\
$^4$ National High Magnetic Field Laboratory, Tallahassee, Florida 32310, USA\\
$^5$Institute of Physics, Slovak Academy of Sciences,
Bratislava, Slovak Republic\\
$^6$College of Optical Sciences, University of Arizona,
Tucson, Arizona 85721, USA\\
$^7$ Department of Physics, Virginia Polytechnic Institute and State
University, Blacksburg, VA 24061, USA\\
$^8$ Department of Physics and Astronomy and HPC\,$^2$ Center for Computational 
Sciences, Mississippi State University, Mississippi 39762-5167, USA
}
\date{\today}

\begin{abstract}
The nanoscopic structure and the stationary propagation velocity of
$(1+1)$-dimensional solid-on-solid interfaces in an Ising lattice-gas
model, which are driven far from
equilibrium by an applied force, such as a magnetic field or a
difference in (electro)chemical potential, are studied by an analytic
nonlinear-response approximation [P.~A.\ Rikvold and M.~Kolesik, J.\ Stat.\
Phys.\ {\bf 100}, 377 (2000)] together with
kinetic Monte Carlo simulations. Here we consider the case 
that the system is coupled to  a two-dimensional phonon bath. In the 
resulting dynamic 
[K.~Saito, S.~Takesue, and S.~Miyashita, Phys.\ Rev.\ E {\bf 61}, 2397 (2000);
K.~Park and M.~A. Novotny, Comput.\ Phys.\ Commun. {\bf 147}, 737 (2002)],
transitions that conserve the system energy are forbidden, and 
the effects of the applied force and the interaction
energies do not factorize (a so-called hard dynamic). In full agreement 
with previous general theoretical results we
find that the local interface width  changes dramatically
with the applied force. 
However, in contrast with other hard dynamics, this change is
nonmonotonic in the driving force. 
Results are also obtained for the
force-dependence and anisotropy of the interface velocity, which also
show differences in good agreement with the theoretical expectations for
the differences between soft and hard dynamics. 
However, significant differences between theory and simulation 
are found near two special values of the driving force, where certain
transitions allowed by the solid-on-solid model become forbidden by the
phonon-assisted dynamic. 
Our results show that different stochastic interface dynamics
that all obey detailed balance and the same conservation laws
nevertheless can lead to radically different interface
responses to an applied force. 
Thus they represent a significant step toward providing a solid physical
foundation for kinetic Monte Carlo simulations. 
\end{abstract}

\pacs{ 
68.35.Ct 
75.60.Jk 
68.43.Hn 
05.10.Ln 
}

\maketitle


\section{Introduction}
\label{sec:INTRO}

Moving internal boundaries or interfaces separating different regions are present in many problems in nature, and the challenge of understanding the dynamics of 
such processes has become increasingly important. In recent years 
considerable efforts have been made toward understanding the large-scale 
structures of growing interfaces. \cite{BARA95,MEAK98} 
In contrast, there has been little 
work related to the microscopic and nanoscopic scales. This is surprising 
since the nanoscopic interface structure plays a crucial role in important
interface properties such as mobility and catalytic and chemical activity.
Technologically, as the sizes of the smallest man-made structures decrease, 
interfacial properties become essential and even dominant. 
Nanoscale assemblies with highly ordered building blocks, such as 
quantum dots\cite{THOR05,MEUN06}
and quantum wires, must be fabricated on a surface or through an interface.

The basic mechanisms of interface growth are complex and often unknown.
A standard way to deal with this problem is constructing a stochastic
model that reproduces essential features. However, extreme care has to
be taken with this approach. Recent studies indicate that different
stochastic dynamics, even when they have the same conserved quantities 
and satisfy
detailed balance, lead to important differences in the nanostructure of
field-driven interfaces.\cite{RIKV00B,RIKV02,RIKV02B,RIKV03,BUEN06}
Surfaces driven by hard dynamics (in which the single-site transition
rates cannot be factorized into one term that depends only on the
interaction energies and a second term that depends only on the field
energies, in contrast with soft dynamics for which this factorization is
possible\cite{MARR99}), such as Glauber, Metropolis, and the two-step
transition dynamics approximation (TDA),\cite{ALAN92,ALAN92A} have a
strong dependence on the applied field. For all hard dynamics studied so
far, the average step
height increases dramatically with increasing field. In contrast, interfaces
driven by soft dynamics, such as the soft Glauber\cite{RIKV02} and the
one-step-dynamics (OSD),\cite{BUEN06,KANG89,FICH91} 
are at most only weakly dependent
on the field and relatively smooth. Furthermore, interfaces driven
by hard dynamics, such as Glauber\cite{RIKV02B} and TDA,\cite{BUEN06}
display significant asymmetry 
between the spin populations on their leading and trailing edges, 
while interfaces moving under soft dynamics either display no
(soft Glauber\cite{RIKV02}) or only weak (OSD\cite{BUEN06}) anisotropy. 

In this paper we study by kinetic Monte Carlo (MC) simulation and a
dynamic mean-field approximation 
the motion of a Burton-Cabrera-Frank solid-on-solid (SOS) interface 
\cite{BURT51} that evolves under a non-conservative dynamic resulting from 
coupling the system to a phonon heat bath. SOS interfaces belong to the 
Kardar-Parisi-Zhang 
(KPZ) dynamic universality class,\cite{BARA95,KARD86} in which 
the macroscopic, stationary distribution for
moving interfaces is Gaussian, corresponding to a random walk with independent
increments. The phonon-assisted dynamic is obtained by introducing a 
weak, linear coupling between a
square-lattice Ising quantum ferromagnet and a phonon (i.e., bosonic)
heat bath attached to the spin system. The transition rates have been calculated using the quantum-mechanical
density matrix equation \cite{SAIT00,PARK02A,PARK02B,HENR05,PARK07} 
and most recently also by the lattice-frame method.\cite{SOLO07}
Both methods give consistent results. The resulting dynamic
is quite different from the Glauber dynamic, which can be similarly 
derived from coupling to fermionic baths. \cite{MART77} 
In particular, for phonon baths of dimension greater than one,
 the phonon-assisted dynamic prohibits transitions that
 conserve the system energy, even if they are allowed by the SOS 
 restriction.
As a result, the model becomes non-ergodic
near special values of the driving field, and the interfaces 
can get stuck in metastable states. The average step height and propagation
velocity therefore become nonmonotonic functions of the field. 
Phonon-assisted dynamics are
relevant in a great variety of physical phenomena, ranging from the 
non-linear optical response of semiconductors \cite{AXT96} to the dynamics 
of quantum dots.\cite{THOR05,MEUN06}
The derivation of a phonon-assisted stochastic dynamic is a significant
step toward putting kinetic MC on a solid physical foundation. 

In this paper we derive analytic, approximate expressions for the interface 
propagation velocity as a function of field, temperature, and interface 
orientation. Our approach is based on a mean-field approximation that 
assumes that individual steps on the interface are statistically 
independent, \cite{RIKV02,RIKV02B,RIKV03} i.e., short-range correlations 
are neglected. This limitation becomes apparent when 
we compare the analytical and simulated results.

The remainder of this paper is organized as follows. 
In Sec.~\ref{sec:MODEL} we introduce the SOS interface model and give the
transition rates for the phonon-assisted dynamic. 
Also in this section we summarize the mean-field approximation for the time 
evolution of the single-step probability density function (pdf), 
as well as its stationary form. We further give 
expressions for the spin-class populations and
interface velocity in terms of the applied field, the temperature, and
the angle of the interface relative to 
the lattice axes.
In Sec.~\ref{sec:MC} we compare simulations and analytical predictions 
for the detailed stationary interfacial nanostructure, including the asymmetry
of the simulated nonequilibrium interfaces. 
A summary and conclusions are provided in Sec.~\ref{sec:DISC}.

\section{Model and Dynamics}
\label{sec:MODEL}

The SOS interfaces are described by the nearest-neighbor 
$S=1/2$ Ising Hamiltonian with anisotropic, ferromagnetic interactions $J_x$ and $J_y$ in the $x$ and $y$ direction, 
respectively: 
\begin{equation}
{\cal H} = -\sum_{x,y} s_{x,y} \left( J_x s_{x+1,y} + J_y s_{x,y+1} 
+ H \right) 
\;, 
\label{eq:ham}
\end{equation}
where $s_{x,y}=\pm1$, $\sum_{x,y}$ runs over all sites, and 
 the applied field $H$ is the driving force. The interface is introduced by fixing 
$s_{x,y}=+1$ and $-$1 for large negative and positive $y$, respectively.  
Without loss of generality we take $H \ge 0$, such 
that the interface on average moves in the positive $y$ direction. 
This Ising model is equivalent to a lattice-gas model with local
occupation variables $c_{x,y} \in \{0,1\}$.\cite{YANG52B,PATH96}
Specifically, we identify $s=+1$ with $c=1$ (occupied or ``solid'') and
$s=-1$ with $c=0$ (empty or ``fluid''). 

\begin{figure}[ht] 
\includegraphics[angle=0,width=.50\textwidth]{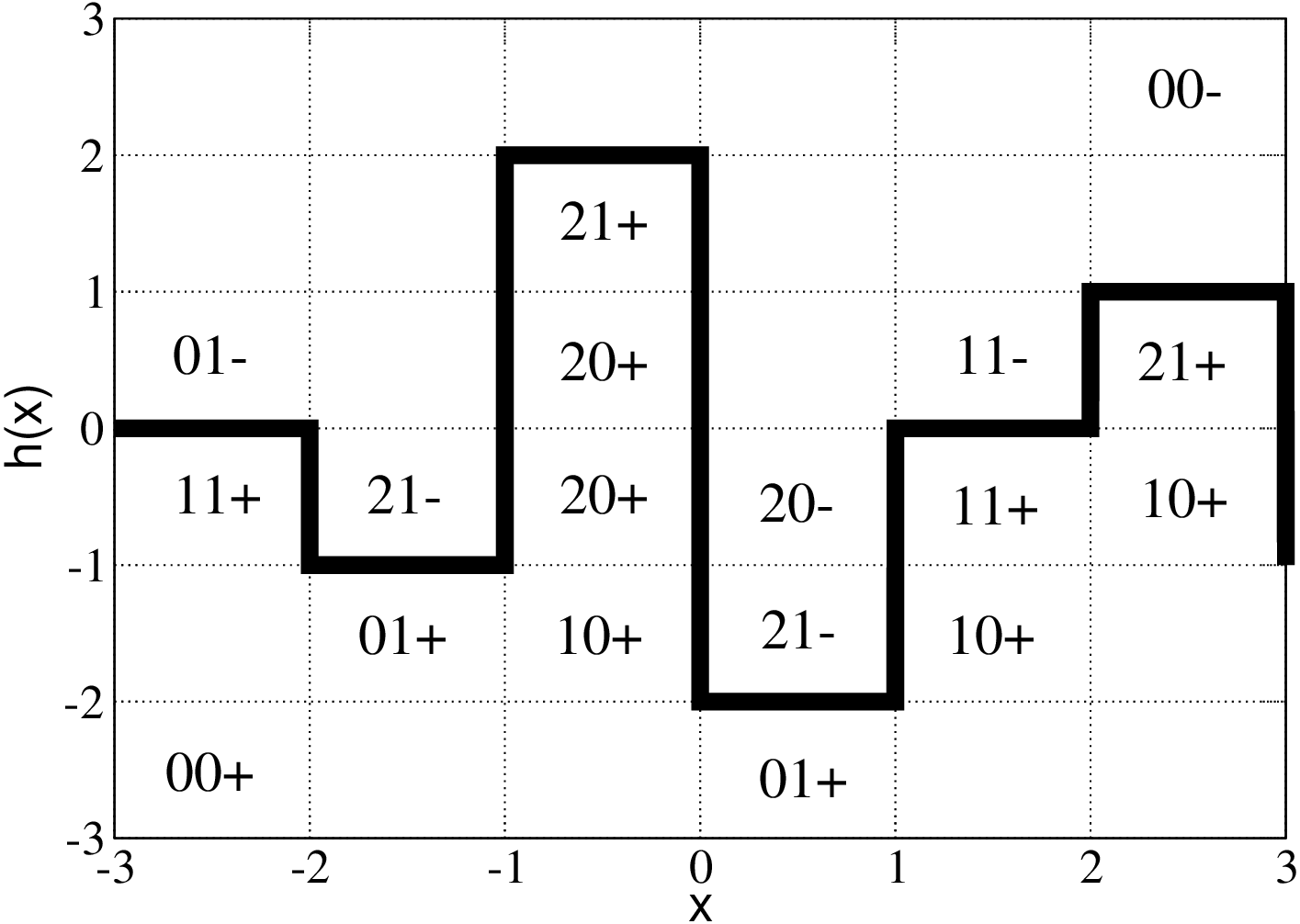}. 
\caption[]{
A short segment of an SOS interface $y=h(x)$ between a positively
magnetized phase (or ``solid'' phase in the lattice-gas picture) 
below and a negative (or ``fluid'') phase
above. The step heights are $\delta(x) = h(x+1/2) - h(x-1/2)$. Interface
sites representative of the different SOS spin classes (see
Table~\protect\ref{table:class} 
) are marked with the
notation $jks$ explained in the text. Sites in the uniform bulk phases are
$00-$ and $00+$. This interface was generated with a symmetric step-height 
distribution, corresponding to $\phi = 0$. 
}
\label{fig:pict}
\end{figure}

The SOS model considers an interface in a lattice gas or $S=1/2$
Ising system on a square lattice of unit lattice constant
as a single-valued integer function $h(x)$ of the $x$-coordinate, with
steps $\delta(x) = h(x+1/2) - h(x-1/2)$ at integer values of $x$.
A typical SOS interface configuration is shown in Fig.~\ref{fig:pict}. In
this paper the two possible states of the site
$(x,y)$ are denoted by the two Ising spin values $s_{x,y} = \pm1$.
(In order that the step positions and the interface heights be
integer as stated above, we place the spins at 
half-integer values of $x$ and $y$, i.e., at the centers of the
unit cells separated by dotted lines in Fig.~\ref{fig:pict}.)
                                                                          
The interface is made to evolve under the phonon-assisted dynamic, a
single-spin-flip (nonconservative) set of transition rates that satisfy detailed 
balance for the allowed transitions.
In most cases this ensures the approach to
equilibrium, which in this case is a uniformly positive
phase with the interface pushed off to positive infinity. 
(For exceptions, see below.) 
The dynamic is defined by the single-spin transition rates, 
$W(s_{x,y} \rightarrow -s_{x,y}) = W(\beta \Delta E)$. Here
$\beta$ is the inverse of the temperature $T$ (Boltzmann's constant is
taken as unity), and $\Delta E$ is the energy change corresponding to a
successful spin flip.
The detailed-balance condition (valid for transitions between
allowed states) is expressed as
$W(\beta \Delta E) / W(- \beta \Delta E)
= e^{- \beta \Delta E}$.

\begin{figure}[ht] 
\vspace{1.0truecm}
\begin{center}
\includegraphics[angle=0,width=.47\textwidth]{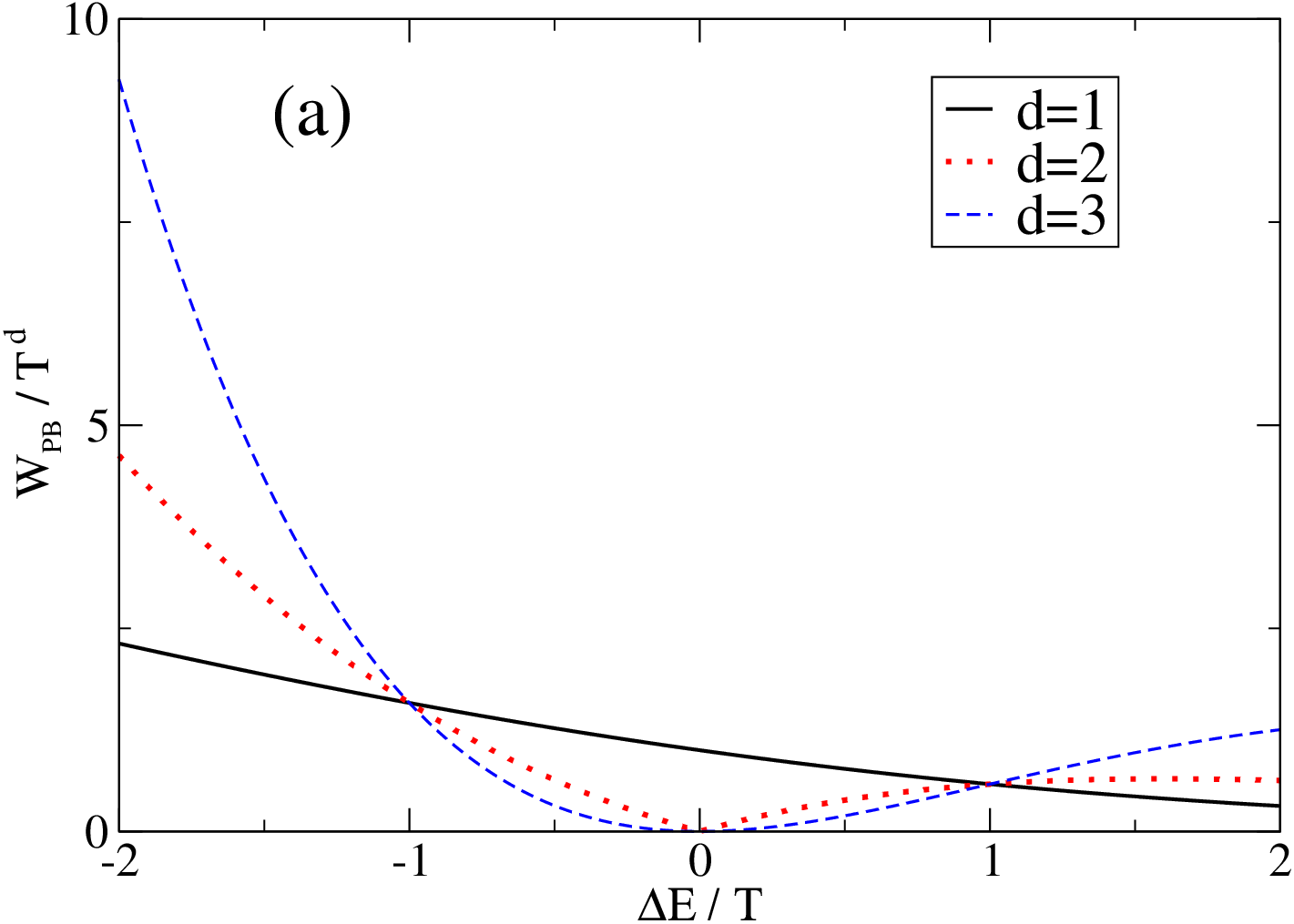}
\hspace{0.5truecm}
\includegraphics[angle=0,width=.47\textwidth]{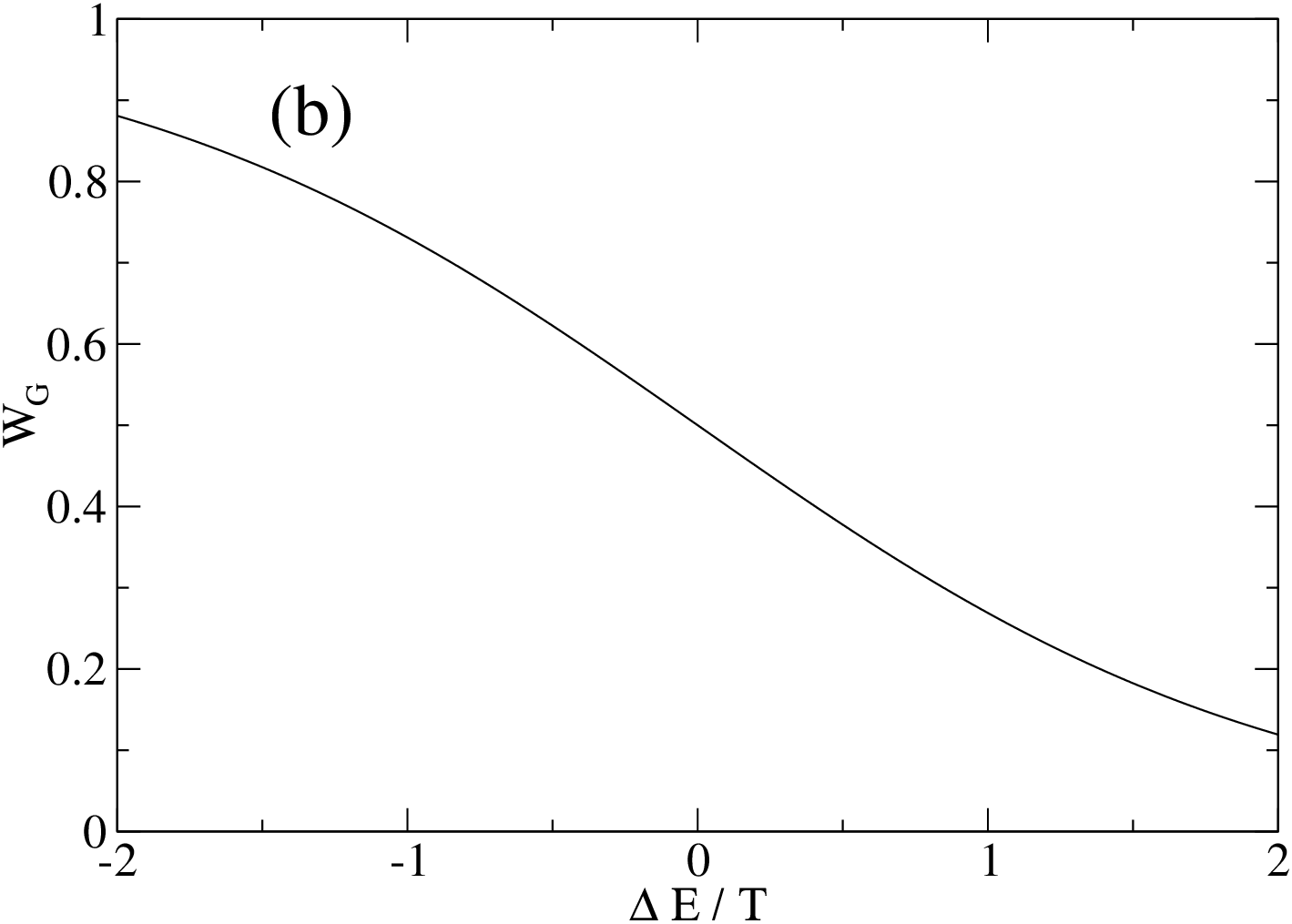}
\end{center}
\caption[]{
(a) (Color online) The transition rates for the $d$-dimensional 
phonon-assisted dynamic,
$W_{\rm PB}$, shown scaled by $T^d$ vs the energy difference 
$\Delta E$ scaled by $T$. 
(b) The transition rates for the standard Glauber dynamic, $W_{\rm G}$,
shown vs $\Delta E/T$.
}
\label{fig:transitions}
\end{figure}

The transition rates for the phonon-assisted
dynamic are defined as \cite{PARK02A,PARK02B}
\begin{equation}
W_{\rm {PB}}(T,\Delta E)=\left| \frac{\Delta E^d}{e ^{\beta\Delta E}-1} \right|,
\label{eq:wpb}
\end{equation}
where $d \in \lbrace 1,2,3 \rbrace$ is the dimension of the 
bosonic heat bath. 
Physically, this rate is the product of three factors: the phonon
occupation number $(e^{\beta |\Delta E|} -1)^{-1}$,
the phonon density of states, proportional to $|\Delta E|^{d-1}$, 
and the magneto-elastic spin-phonon coupling, proportional to 
$|\Delta E|$.\cite{PARK02A,SOLO07}
For $d=2$ and $3$, $W_{\rm{PB}}(T,\Delta E=0)=0$, while for $d=1$ the transition
rate is nonzero and smooth at $\Delta E =0$. 
These transition rates are plotted in Fig.~\ref{fig:transitions}(a). For 
comparison we also plot the transition rates for the Glauber dynamic 
in Fig.~\ref{fig:transitions}(b). In the present work, with the exception of 
Fig.~\ref{fig:vd1} we use $d=2$, i.e, a two-dimensional heat bath. 
Thus, the transition rates vanish linearly with $|\Delta E|$ near $|\Delta E|=0$.
It should, however, be emphasized that the derivations of
Eq.~(\ref{eq:wpb}) are based on a {\it weak, linear\/} coupling of the phonon
bath to the spin system. It is therefore possible that nonlinear 
and/or multiphonon effects
may set a lower bound on physical transition rates for $\Delta E$ near zero, 
and thus restore the ergodicity of the spin model. 
However, we note that recent experiments on phonon-mediated spin
relaxation in a quantum dot shows a significant decrease in the relaxation
rate for transitions involving $\Delta E$ near zero.\cite{MEUN06}

Notice that the phonon-assisted transition rates cannot be factorized into one 
part that depends only on
the interaction energy and another that depends only on the applied field;
thus it belongs to the class of dynamics defined as hard.\cite{RIKV02,MARR99}   
In order to preserve the
SOS configuration at all times, flips are allowed only at
sites which have exactly one broken bond in the $y$ direction.

With the Ising Hamiltonian, there are only a finite 
number of different values of $\Delta E$. The spins can therefore be 
divided into ten classes,\cite{SPOH93,GILM76,BORT75,NOVO95A} 
labeled by the spin value $s$ and the number 
of broken bonds between the spin and its nearest neighbors in the $x$ and 
$y$ directions, $j$ and $k$, respectively. 
The spin classes consistent with the SOS model are denoted $jks$ 
with $j \in \{0,1,2\}$ and $k \in \{0,1\}$. They are shown in
Fig.~\ref{fig:pict} and listed in 
Table~\ref{table:class}. At $H=0$, $\Delta E=0$ for transitions between
$11-$ and $11+$ (diffusion of steps of unit height). 
Thus these transitions are forbidden for $d=2$ and $3$. At $H=2J_x$, the transitions forbidden for $d=2$ and $3$ are between $01-$ and $21+$ (nucleation or elimination of a knob of stable (+) 
phase on a smooth, horizontal interface). 
For other values of (nonnegative) $H$, no transitions allowed by the 
SOS condition are forbidden.

In the SOS model and our analytical approximation the heights of the individual 
steps are assumed to be statistically independent and 
identically distributed. This assumption is exact for $H=0$.\cite{BURT51} 
The step-height probability density function (pdf) 
is given by the interaction energy corresponding to the $|\delta(x)|$ broken 
$J_x$-bonds between spins in the columns centered 
at $(x-1/2)$ and  $(x+1/2)$ as  
\begin{equation}
p[\delta(x)] = Z(\phi)^{-1} X^{|\delta(x)|}
\ e^{ \gamma(\phi) \delta(x) } \;. 
\label{eq:step_pdf}
\end{equation}
The factor $X$ determines the width of the pdf, 
and $\gamma(\phi)$ is a Lagrange multiplier which maintains the mean step 
height at an $x$-independent value, $\langle \delta(x) \rangle = \tan \phi$, 
where $\phi$ is the overall angle between the interface and the $x$ axis. 
$Z(\phi)$ is a partition function that will be discussed below. 
In equilibrium, $X$ is simply the Boltzmann 
factor, $e^{- 2 \beta J_x}$, which is independent of $H$. In previous papers
\cite{RIKV00B,RIKV02B} an expression for a field-dependent $X(T,H)$ was
obtained, based on a dynamic mean-field approximation for the equation of motion for the
single-step pdf together with a detailed-balance argument for the stationary state. This improved
non-linear response approximation gives (see Ref.~\onlinecite{RIKV02B} for details of the calculation),
\begin{equation}
X(T,H) = e^{-2 \beta J_x} 
\left\{
\frac{e^{-2 \beta H}W[\beta(-2H-4J_x)] + e^{2\beta H}W[\beta(2H-4J_x)]}
{W[\beta(-2H-4J_x)] + W[\beta(2H-4J_x)]}
\right\}^{1/2}
\;,
\label{eq:XTH}
\end{equation}
which is independent of $\gamma(\phi)$. The dependence on the specific dynamic is evident here by the presence of the transition rates associated with the reversal of a single spin, $W(\beta \Delta E)$. For $H=0$, $X$ is reduced to its equilibrium value, $X(T,0) = e^{-2 \beta J_x}$. For soft dynamics, where the field and the interaction terms factorize, the $H$-dependence in Eq.~(\ref{eq:XTH})
cancels out, while for hard dynamics $X$ has a nontrivial dependence on $H$. 
Early results indicated that the SOS interfaces generated with the soft 
Glauber dynamic are indeed independent of $H$.\cite{RIKV02} However
interfaces generated with the OSD dynamic, which is also soft, show a weak  
dependence on the interface structure of the field.\cite{BUEN06}

The partition function for the interface is 
\begin{equation}
Z(\phi)\begin{large}\end{large}
=
\sum_{\delta = -\infty}^{+\infty} X^{|\delta|} e^{ \gamma(\phi) \delta } 
= 
\frac{1-X^2}{1 - 2 X \cosh \gamma(\phi) + X^2} ,
\label{eq:Z}
\end{equation}
where $\gamma(\phi)$ is given by 
\begin{equation}
e^{\gamma (\phi)} 
= 
\frac{ \left(1+X^2 \right)\tan \phi 
+ \left[ \left( 1 - X^2 \right)^2 \tan^2 \phi + 4 X^2 \right]^{1/2}}
{2 X \left( 1 + \tan \phi \right)} 
\label{eq:chgam}
\end{equation}
(see details in Refs.~\onlinecite{RIKV00B} and \onlinecite{RIKV02B}).

The mean spin-class 
populations, $\langle n(jks) \rangle$, are all obtained from the 
product of the independent pdfs for $\delta(x)$ and $\delta(x$+1). 
Symmetry of $p[\delta(x)]$ under the transformation 
$(x,\phi,\delta) \rightarrow (-x,-\phi,-\delta)$ ensures that 
$\langle n(jk-) \rangle = \langle n(jk+) \rangle$ for all $j$ and $k$. 
Numerical results illustrating the breakdown of this up-down symmetry for 
large $H$ are discussed in Sec.~\ref{sec:MC}. The general expressions for
the class populations are given in the third column of Table~\ref{table:class}; details
of the calculation can be found in Ref.~\onlinecite{RIKV02B}.

Whenever a spin  flips from $-1$ to $+1$, 
the corresponding column of the interface advances by one lattice constant 
in the $y$ direction. Conversely, the column 
recedes by one lattice constant when a spin 
flips from +1 to $-$1. The corresponding energy changes are 
given in the second column in Table~\ref{table:class}. 
Since the spin-class populations on both sides of the 
interface are equal in this approximation, the contribution 
to the mean velocity in the $y$ direction 
from sites in the classes $jk-$ and $jk+$ becomes 
\begin{equation}
\langle v_y(jk) \rangle 
= 
W \left( \beta \Delta E(jk-) \right)
-
W \left( \beta \Delta E(jk+) \right) 
 \;. 
\label{eq:generalv}
\end{equation}
The mean propagation velocity perpendicular to the interface becomes  
\begin{equation}
\langle v_\perp (T,H,\phi) \rangle 
= 
\cos \phi \sum_{j,k} \langle n(jks) \rangle \langle v_y (jk) \rangle 
\;, 
\label{eq:totalv}
\end{equation}
where the sum runs over the classes included in Table~\ref{table:class}. 
It was shown in Ref.~\onlinecite{RIKV02B} 
that Eq.~(\ref{eq:totalv}) reduces to the results for 
the single-step\cite{DEVI92,SPOH93,MEAK86,PLIS87} and the polynuclear 
growth\cite{DEVI92,KRUG89,KERT89}
models at low temperatures for large and small $\phi$, respectively. 
The spin-class populations 
  listed in Table~\ref{table:class} 
can be calculated explicitly by replacing $X$ with its
corresponding value from Eq.~(\ref{eq:XTH}). 

In the next Section we show that the nonlinear-response approximation 
gives good agreement with MC simulations of driven SOS interfaces 
evolving under the phonon-assisted dynamic for a wide range of fields and
temperatures. The main deviations between theory and simulations are seen 
for $H/J=0$ and $2$, where some transitions allowed by the SOS 
restrictions have $\Delta E =0$ and thus are forbidden by the 
two-dimensional phonon-assisted dynamic, Eq.~(\ref{eq:wpb}).

\section{Comparison with Monte Carlo Simulations} 
\label{sec:MC} 

We calculated the step-height distributions, propagation velocities, 
and spin-class populations, analytically and by kinetic MC simulations,
for the phonon-assisted dynamic in the isotropic case, $J_x = J_y = J$. 
The details of our particular implementation 
of the $n$-fold way rejection-free MC 
algorithm\cite{BORT75, GILM76} are described in 
Refs.~\onlinecite{RIKV00B} and \onlinecite{RIKV02B}. The extension to
continuous time, which is necessary to handle transition rates greater
than unity, was introduced in Ref.~\onlinecite{BUEN06}. 

The numerical results presented here are based on 
MC simulations mostly at the two 
temperatures, $T = 0.2T_c$ and~0.6$T_c$ ($T_c = -2J/\ln(\sqrt{2} -1)
\approx 2.269J$ is the critical temperature for the isotropic,
square-lattice Ising model\cite{ONSA44}), with  
$L_x = 10\,000$ and fixed $\phi$ between 0 and $45^\circ$. 
In order to ensure stationarity we ran the simulation for $10^4$ 
$n$-fold way updates per updatable spin (UPS)  
for thermalization before taking any measurements.
Unless otherwise noted, the initial condition before thermalization was
a microscopically flat interface. The initial condition only makes a
difference near $H=0$. 
Stationary 
class populations and interface velocities were averaged over $10^6$~UPS. 
For the stronger fields at $T=0.2T_c$ we used ten times as many UPS. 
Adequate statistics for the step-height pdfs were ensured by the large $L_x$.

\subsection{Stationary single-step probability densities}
\label{sec:MCss}

\begin{figure}[ht] 
\begin{center}
\includegraphics[angle=0,width=.47\textwidth]{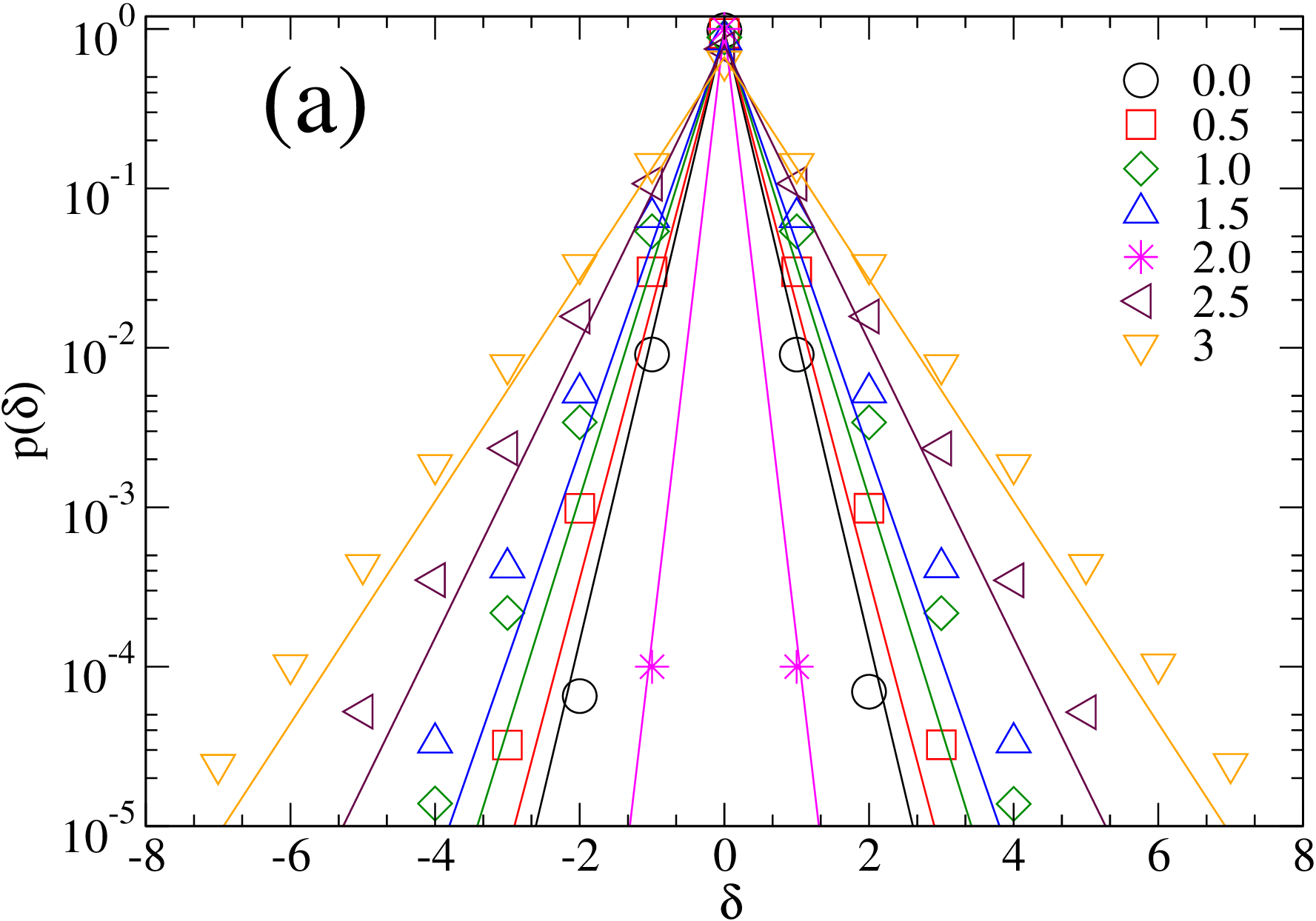}
\hspace{0.5truecm}
\includegraphics[angle=0,width=.47\textwidth]{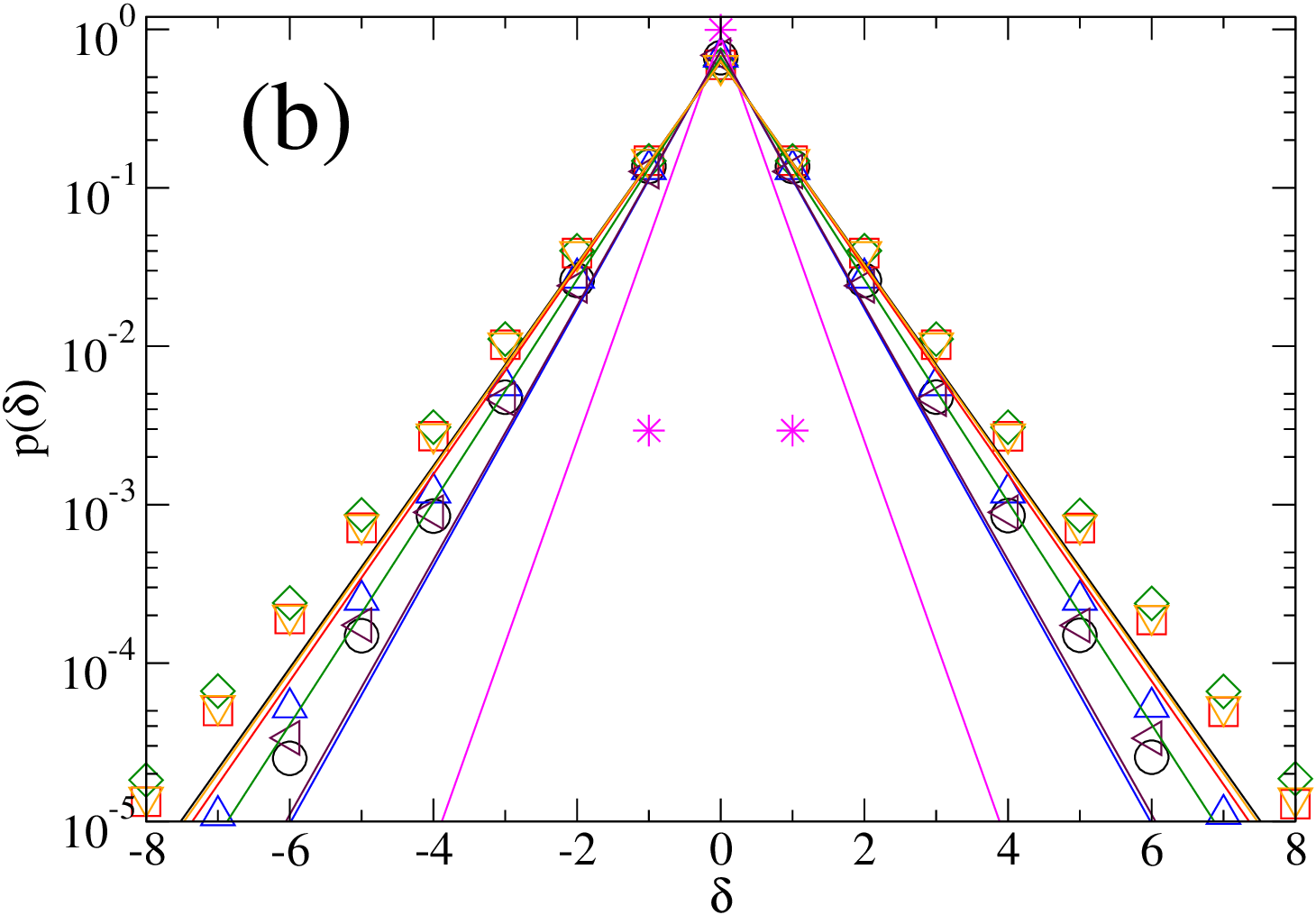}
\end{center}
\caption[]{
(Color online) MC (data points) and analytical (solid lines) 
results for the stationary 
single-step pdf, shown on a logarithmic
 scale vs $\delta$, for the values of $H/J$ given in the legend.
(a) $T=0.2T_c$. 
(b) $T=0.6T_c$. The symbols (and colors)  have
the same interpretations in (a) and (b). Note the nonmonotonic field dependence near $H/J=2$.
}
\label{fig:pdelta}
\end{figure}

Stationary single-step pdfs were obtained by MC simulation at $T=0.2T_c$ and 
$0.6T_c$ for $\phi = 0$ and 
several values of $H$. The simulation data
and the theoretical results for $p[\delta]$ are shown in Fig.~\ref{fig:pdelta}.
 The theoretical results are calculated
with 
Eq.~(\ref{eq:step_pdf}), with $X(T,H)$ from Eq.~(\ref{eq:XTH}). The agreement between theoretical and simulated results is quite good, particularly at the lower temperature.

\begin{figure}[ht] 
\vspace{1.0truecm}
\includegraphics[angle=0, width=.50\textwidth]{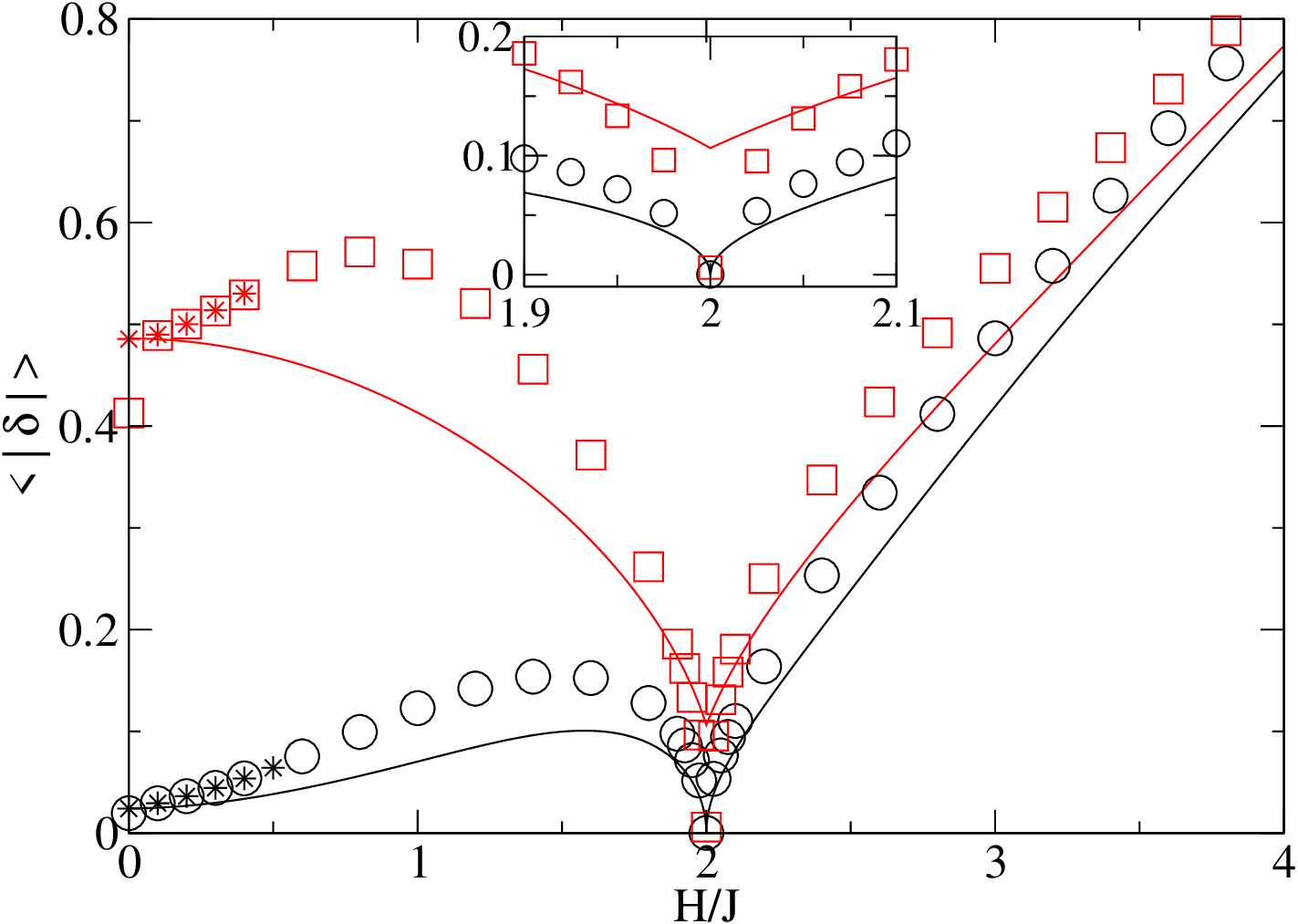}\\
\vspace{1truecm} 
\caption[]{
(Color online) Average stationary step height 
$\langle | \delta | \rangle$ 
vs $H/J$ for $\phi$=0 at $T$=$0.2T_c$ and~0.6$T_c$.
The curves represent the theoretical results.  
Curve with 
circles (black): $T=0.2T_c$. 
Curve with 
squares (gray, red online): $T=0.6T_c$. 
These data refer to interfaces started from a microscopically flat
initial state. 
We also include values near $H/J=0$, calculated by the phonon-assisted 
dynamic using as starting 
state the thermalized interface obtained with the standard 
Glauber dynamic. Black asterisks: $T=0.2T_c$. Grey asterisks (red online): 
$T=0.6T_c$. The differences are only evident near $H/J=0$. 
In this and all the following figures, the statistical uncertainty is
much smaller than the symbol size. The inset shows a magnified view of 
the region around $H/J=2$. Note the disagreement between the theoretical 
and the simulation values at $T=0.6T_c$ when $H/J=2$.

}
\label{fig:dh}
\end{figure}

Another way to compare the analytical and simulation results is by calculating
$\langle | \delta | \rangle$ by averaging over  the simulated step-height pdf,
and comparing these values with the theoretical ones obtained from 
Eq.~(\ref{eq:step_pdf}), 
$\langle | \delta | \rangle = 2X/\left( 1-X^2 \right)$,
with $X$ from Eq.~(\ref{eq:XTH}).  
The results are shown in 
Fig.~\ref{fig:dh} for $\phi = 0$ at $T = 0.2T_c$ and~0.6$T_c$,
calculated theoretically (solid lines) and by MC simulation (symbols). 
The agreement between both results is reasonable. However, the
theoretical data present a smoother dependence on the field than  
obtained from the simulation. 

In Fig.~\ref{fig:dh}, the behaviors of $\langle | \delta | \rangle$ near 
$H/J = 0$ and 2 are of particular interest. 
At $H/J=0$ the system should be in equilibrium, and the theoretical
results are exact. \cite{BURT51} 
The discrepancy between theory
and simulation at this field, especially at $T=0.6T_c$, 
therefore means that the system simulated
with the phonon-assisted dynamic, starting from a microscopically flat initial
state, does not equilibrate completely. This is not due to a too short thermalization time. 
Rather, the reason is the aforementioned suppression by Eq.~(\ref{eq:wpb}) near
$H=0$ of transitions between states $11+$ and $11-$, which correspond to 
diffusion of steps along the interface and represent an important
mechanism for equilibration. 
However, when $\langle | \delta | \rangle$, near $H/J=0$ at $T=0.6T_c$, is 
obtained by the phonon-assisted dynamic starting from the 
thermalized interface generated with the standard Glauber dynamic, 
there is excellent agreement with the theoretical result, as can also be 
seen in Fig.~\ref{fig:dh}. 
As $H/J$ is increased from zero, the effect of the initial condition
rapidly vanishes. 

\begin{figure}[ht]
\begin{center}
\includegraphics[angle=0,width=.47\textwidth]{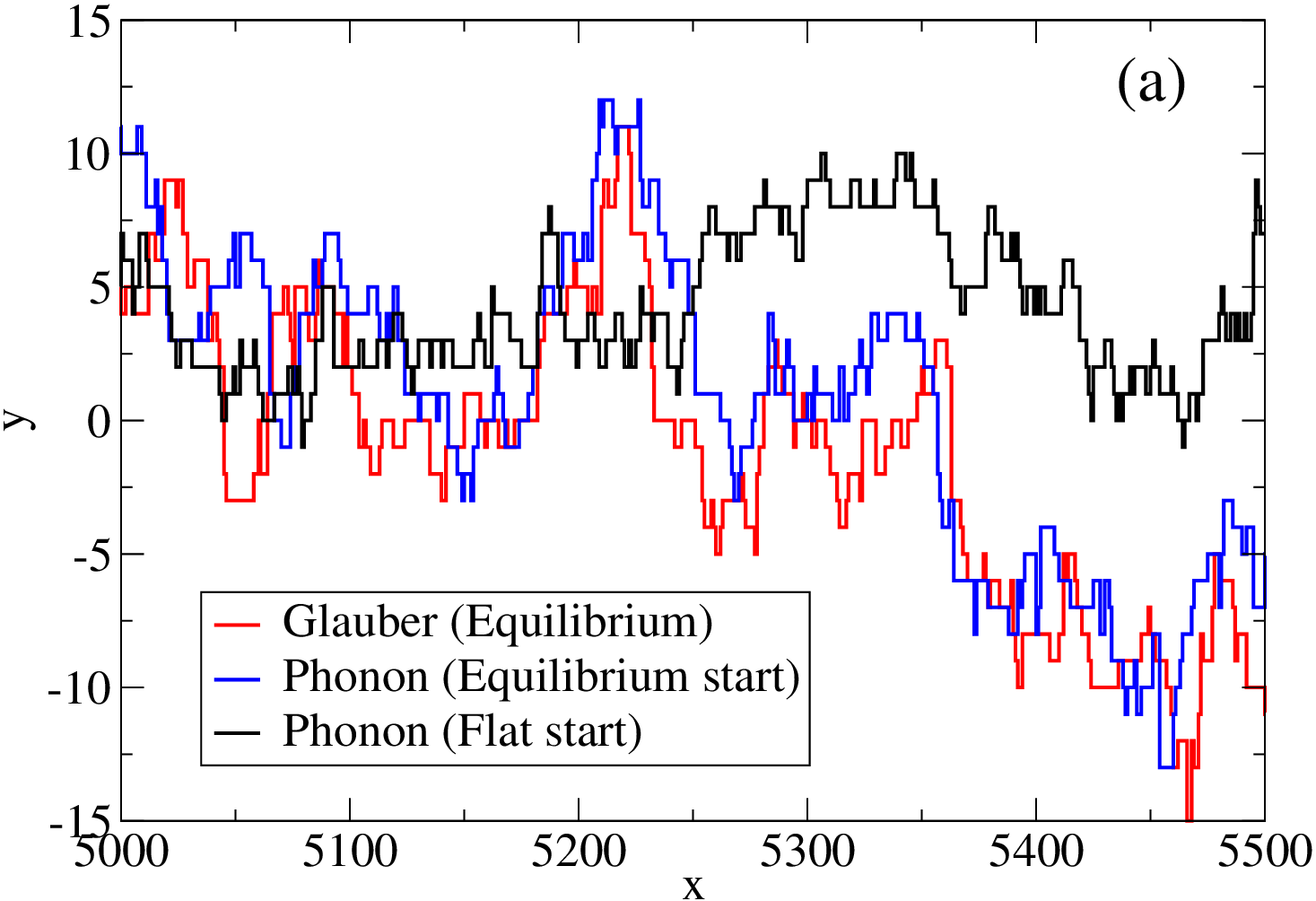}
\hspace{0.5truecm}
\includegraphics[angle=0,width=.47\textwidth]{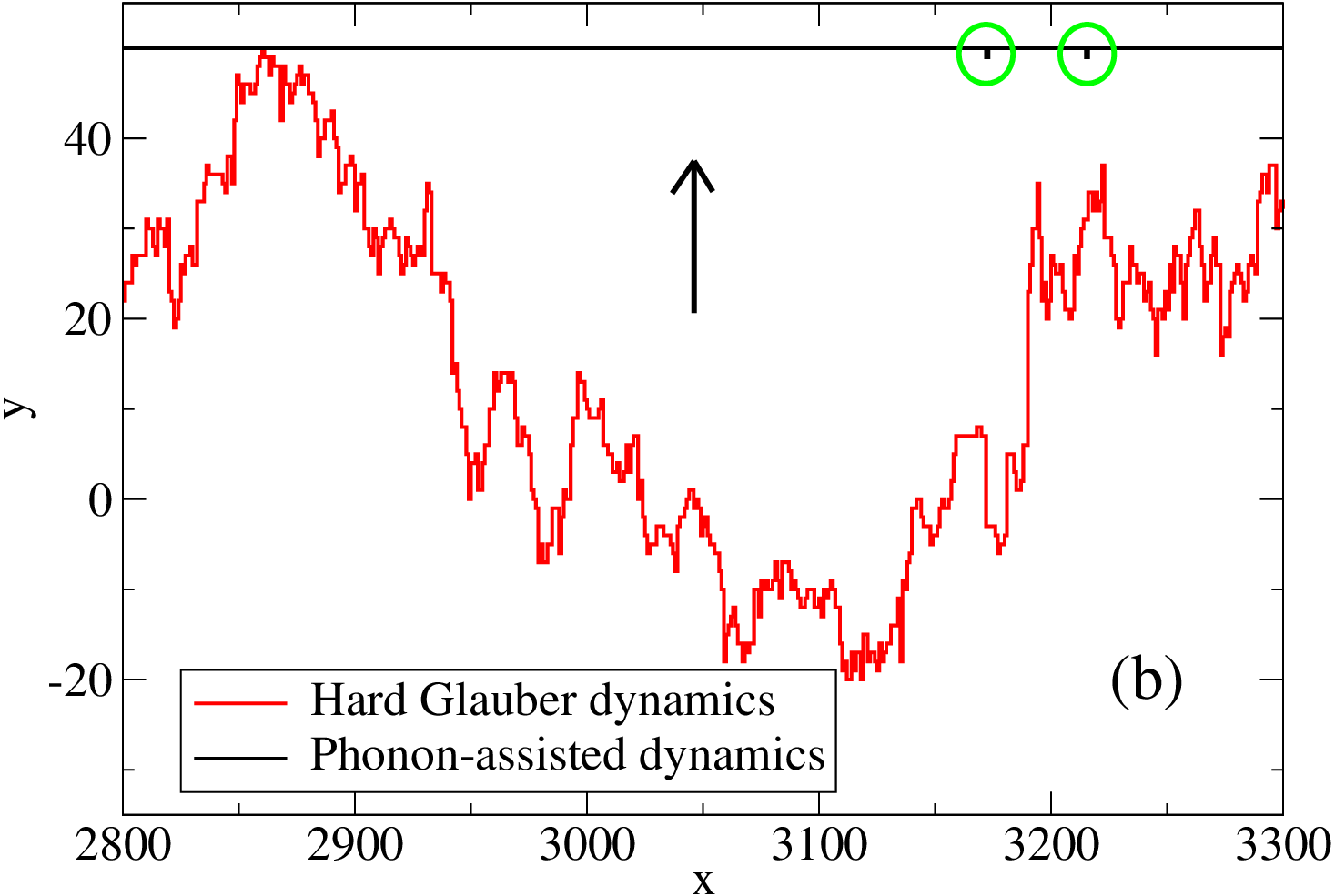}
\end{center}
\caption[]{
(Color online)
Short segments of thermalized interfaces. 
{(a)}
$H/J = 0$, $T=0.6T_c$. 
The three graphs show an equilibrium interface created with the
standard Glauber dynamic over $10^4$~UPS (medium gray, red online); 
an interface
created by the phonon-assisted dynamic over $10^{6}$~UPS, 
using the equilibrium interface as starting state (dark gray, blue online); 
and an interface
created by the phonon-assisted dynamic over $10^{10}$~UPS, 
using a microscopically flat interface as starting state (black). 
{(b)}
$H/J = 2$, $T=0.6T_c$.
The jagged interface (medium gray, red online) is in the statistically
stationary state, propagating in the direction of the arrow under the
standard Glauber dynamic. At a given time, the dynamic is switched to the
phonon-assisted transition rates, using the Glauber interface as initial
state. The lagging parts rapidly catch up with the
absolute maximum of the Glauber interface, where the interface gets permanently
stuck in an almost perfect, microscopically
flat configuration with a very small density of
``backward" $12-$ notches (circled). At this field, the transition 
forbidden by the phonon-assisted dynamic is the nucleation of 
``forward" $12+$ notches,
which are needed to nucleate propagation of a microscopically flat interface. 
}
\label{fig:profile}
\end{figure}

In Fig.~\ref{fig:profile}(a) we show together snapshots of stationary
interfaces at $H/J=0$ for $T=0.6T_c$, generated in three different ways: 
using the standard Glauber dynamic; with the phonon-assisted
dynamic starting from 
the equilibrated interface obtained by the Glauber dynamic; and with the
phonon-assisted
dynamic starting from a microscopically flat interface. Due to the fact 
that energy-conserving moves (horizontal or vertical
step-diffusion) are prohibited by the
phonon-assisted dynamic, the interface started from the equilibrium
interface is highly correlated with the latter, and both have 
$\langle | \delta | \rangle \approx 0.49$, the equilibrium value. 
For the same reason, 
the phonon-assisted interface started from a microscopically
flat interface configuration 
does not fully equilibrate, but settles into a metastable configuration with 
$\langle | \delta | \rangle \approx 0.41$, as seen in Fig.~\ref{fig:dh}.

At $H/J=2$, MC simulations give a value of  
$\langle | \delta | \rangle$ strictly zero for $T=0.2T_c$ and very close to 
zero for $T=0.6T_c$, while the theoretical value is very small but nonzero 
at $T=0.2T_c$, and clearly larger at $T=0.6T_c$.
(See inset in Fig.~\ref{fig:dh}.)
For strong values of the field, the step height is only weakly dependent 
on the temperature.
These results are quite different from those obtained with the 
standard (hard) Glauber dynamic
(see Fig.~5(a) of Ref.~\onlinecite{RIKV02B}). However, the strong $H$ 
dependence of the step heights is characteristic of hard dynamics.

In Fig.~\ref{fig:profile}(b) we show together snapshots of stationary
interfaces at $H/J=2$ and $T=0.6T_c$,  
one thermalized with the standard Glauber dynamic, and the other
generated by the phonon-assisted dynamic, using the Glauber interface
as its starting state. The interface obtained with the phonon-assisted 
dynamic is almost entirely microscopically flat, with a very small density 
of ``backward" ($21-$) notches that are created at a rate 
$\propto$  $\exp(-8J/T)$ and annihilated almost immediately. 
In fact, for $H/J=2$ the interface gets stuck as it can never progress
beyond the absolute maximum of the starting configuration
(and is thus non-ergodic) due to the vanishing rate 
of the transition $01-$ $\rightarrow$ $21+$. One possible way to overcome this situation is to give the 
interface alternative paths to reach equilibrium. This could possibly be done 
by relaxing the SOS constraint to allow overhangs and
bubbles.\cite{RIKV00B,RIKV03}
In contrast, the Glauber interface at the same field and temperature
propagates at a nonzero velocity and is microscopically quite rough, with  
$\langle | \delta | \rangle \approx 1.79$.

\subsection{Stationary interface velocities}
\label{sec:MCi}

\begin{figure}[ht] 
\vspace{1.0truecm} 
\includegraphics[angle=0,width=.50\textwidth]{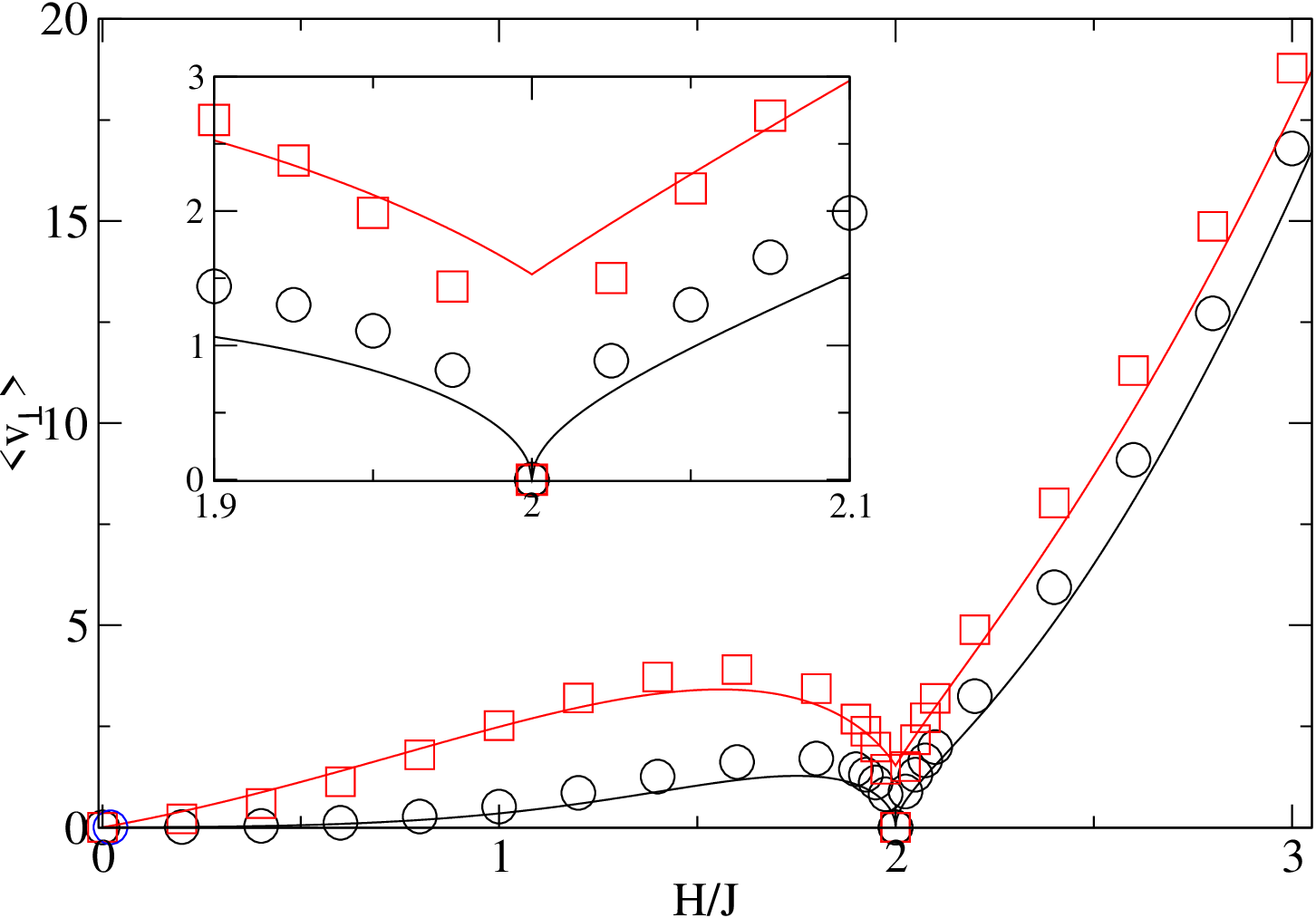}
\caption[]{
(Color online) The average stationary 
normal interface velocity $\langle v_\perp \rangle$ vs $H/J$ for 
$\phi = 0$. The MC results are shown as data points, circles for $T=0.2T_c$ and
squares for $T=0.6T_c$,   
and the theoretical results as solid curves. 
The inset shows a magnified view of the region around $H/J=2$. Note again 
the disagreement between the theoretical and the simulation 
values at $T=0.6T_c$ when $H/J=2$.  
}
\label{fig:vph}
\end{figure}

\begin{figure}[ht]
\begin{center}
\includegraphics[angle=0,width=.47\textwidth]{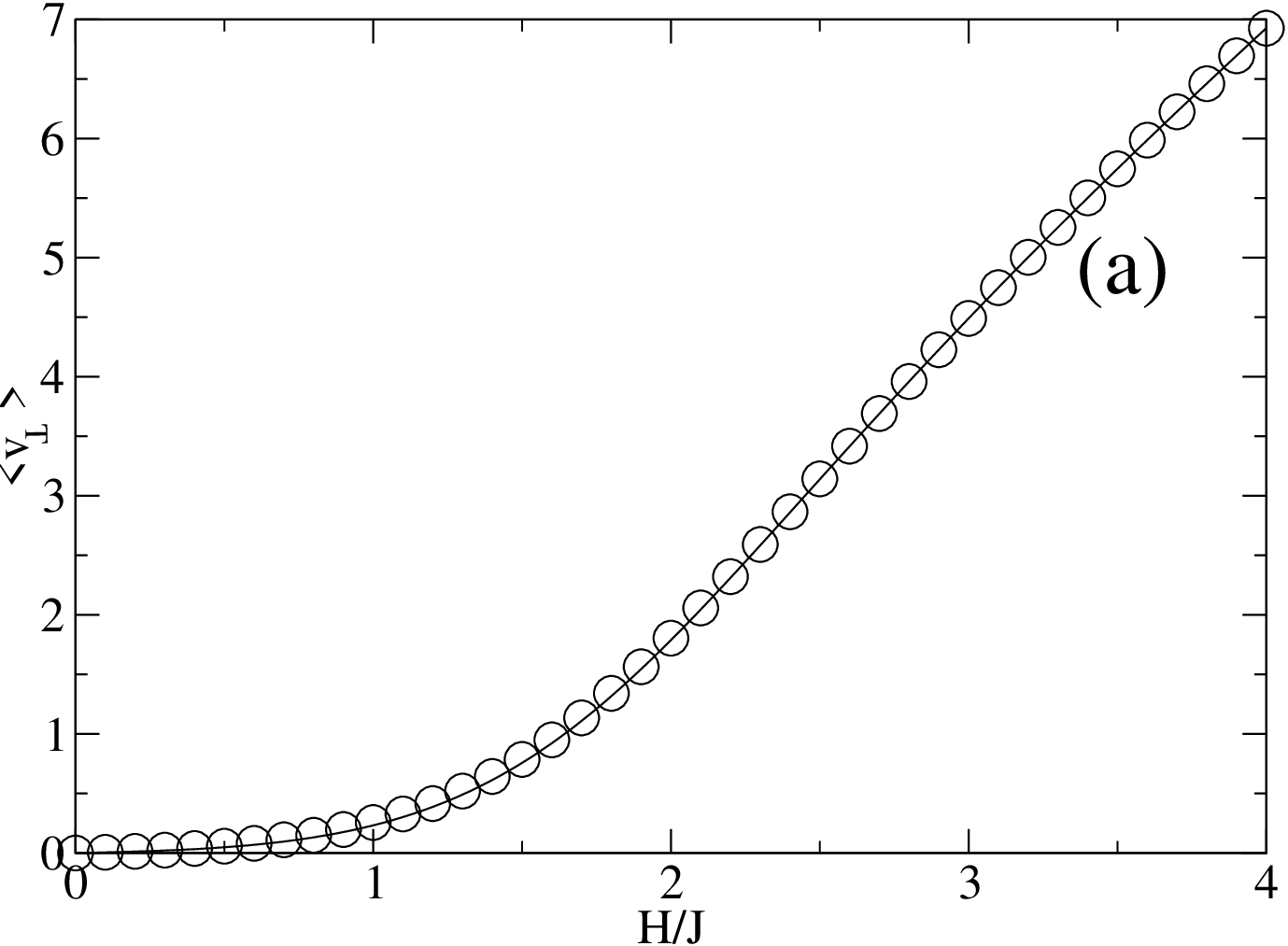}
\hspace{0.5truecm}
\includegraphics[angle=0,width=.47\textwidth]{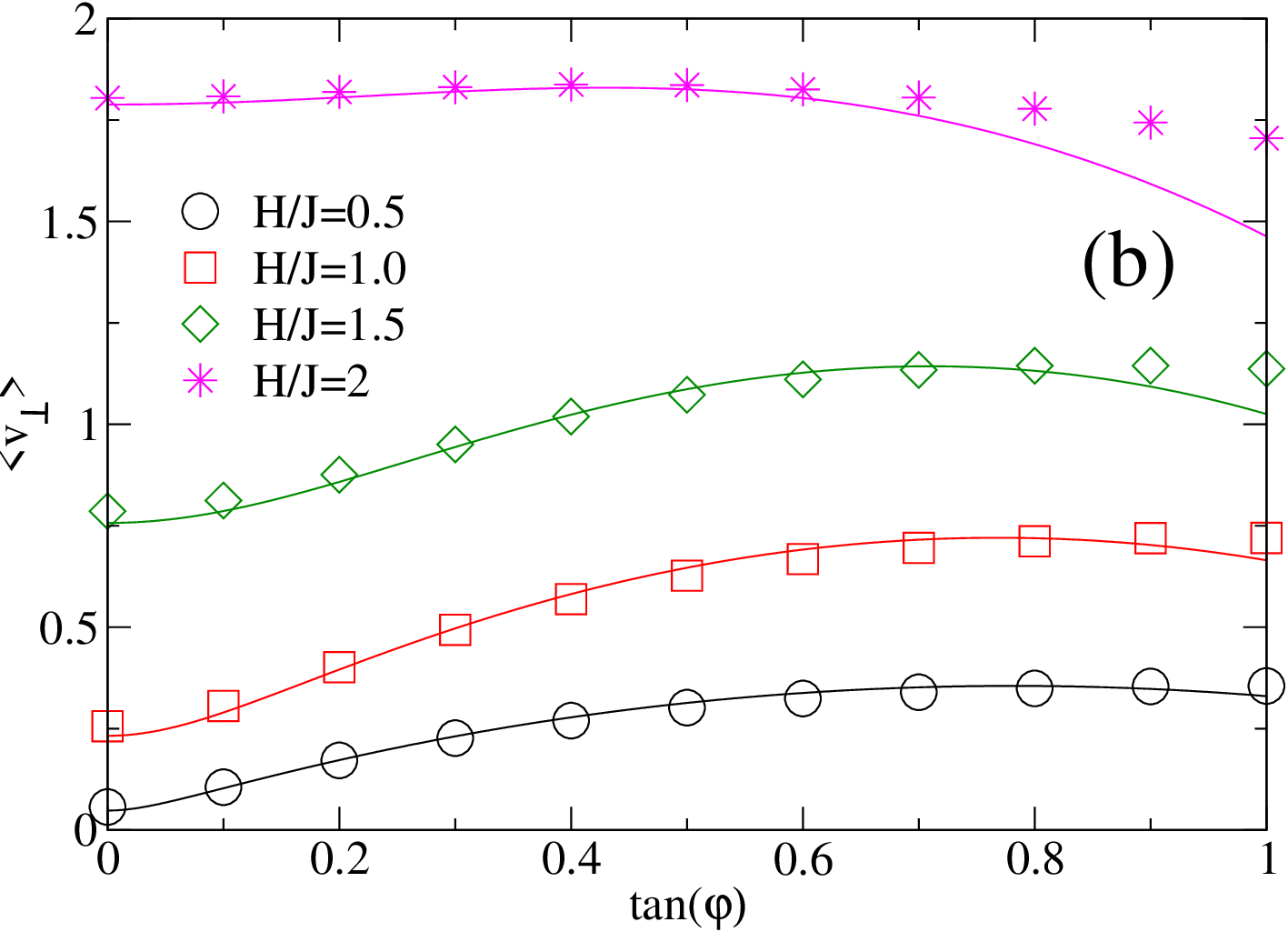}
\end{center}
\caption[]{
(Color online) Average stationary
normal interface velocity at $T=0.2T_c$, obtained by coupling the system to a 
{\it one-dimensional\/} phonon bath. MC data are represented
by the symbols and analytical results by the solid curves. 
(a) Velocity vs $H/J$ for $\phi = 0$. 
(b) Velocity vs $\tan \phi$ for several values of $H/J$. 
}
\label{fig:vd1}
\end{figure}

In Fig.~\ref{fig:vph} we show the mean propagation velocity perpendicular
to the interfaces vs $H/J$ for 
$\phi = 0$, obtained with the analytical approximation, 
Eq.~(\ref{eq:totalv}), and by simulations.
In general there is good agreement between the MC results
and the nonlinear-response theory. However, there is a significant discrepancy 
at $H/J=2$. At this field, the simulated velocities are zero, independent 
of the temperature, while the theoretical value is small but nonzero and 
increases with temperature. This is perfectly consistent with the
microscopically flat interface structure at $H/J=2$, discussed above. 
In fact, the knob-nucleating transition $01- \rightarrow 21+$, which is
forbidden at this field, is precisely the transition needed to nucleate
the advance of a microscopically flat interface. 
When $H/J>2$, the velocity increases rather rapidly with $H$.
This behavior is very different from the one obtained for other hard dynamics 
such as the standard Glauber and the TDA, where the velocity is bounded by 
unity. (See Fig.~6 of Ref.~\onlinecite{RIKV02B} and Fig.~6(a) of 
Ref.~\onlinecite{BUEN06}.)

In contrast to the results discussed above, Fig.~\ref{fig:vd1} shows the 
velocity for the 
case in which the dimension of the phonon bath is unity, i.e., $d=1$, which 
has a transition rate that decreases smoothly and monotonically with 
$\Delta E/T$ (in contrast with the $d=2$ and $3$ cases in which the 
transition rates vanish for $H=0$ and $H/J=2$). The agreement between theory 
and simulation is excellent over the whole range of $H/J$ and tilt angles, except for very large angles at higher fields. (This is also
the case for step-height distributions and other characteristic
quantities.) 

\begin{figure}[ht]
\begin{center}
\includegraphics[width=.47\textwidth,angle=0]{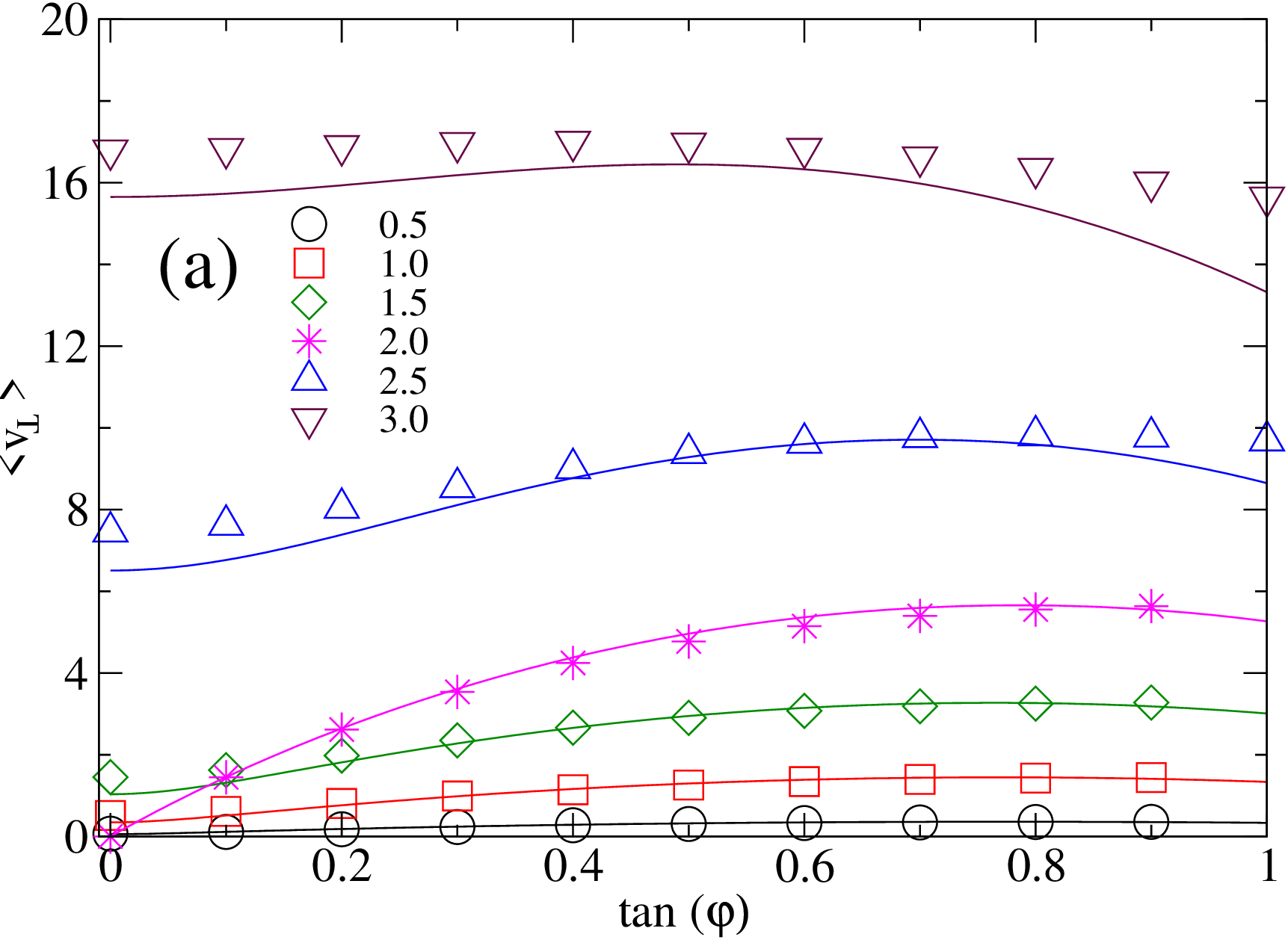}
\hspace{0.5truecm}
\includegraphics[width=.47\textwidth,angle=0]{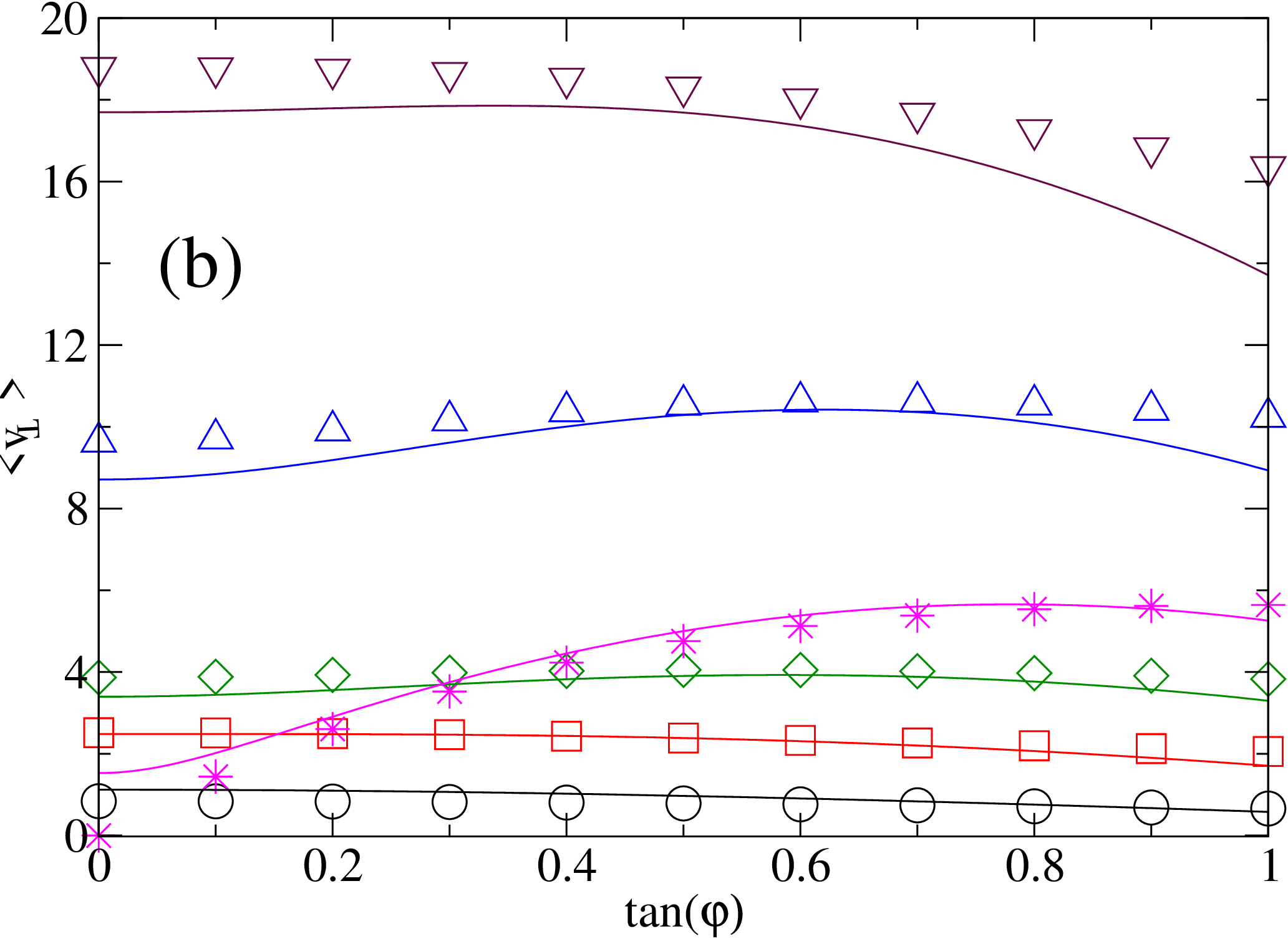}
\end{center}
\caption[]{
(Color online) The average stationary
normal interface velocity $\langle v_\perp \rangle$ vs $\tan \phi$,
for several values of $H/J$. The symbols represent MC data, and the solid curves analytical results.
(a) $T=0.2T_c$,
(b) $T=0.6T_c$. The symbols have the same interpretations in (a) and (b), 
given by the legend in (a). Online, the colors of the curves and symbols match.
}
\label{fig:angle}
\end{figure}

The dependence of the normal velocity on the tilt angle $\phi$ is shown in
Fig.~\ref{fig:angle}(a) and Fig.~\ref{fig:angle}(b) for several values of $H/J$ at $T=0.2T_c$ and $T=0.6T_c$ respectively. The agreement between the theoretical results and the 
simulations is very good except at higher fields,
where the agreement is only good at intermediate values of $\phi$. 
The results are qualitatively similar to those obtained with
other hard dynamics (see Refs.~\onlinecite{RIKV02B} and ~\onlinecite{ BUEN06}). At $T=0.2T_c$, in weak fields the velocity increases with 
$\phi$, in agreement with the polynuclear growth model at 
small angles and the single-step model for larger angles. For strong fields
the behavior changes gradually to the reverse anisotropy of Eden-type
models.\cite{MEAK86B,HIRS86} This is essentially the same behavior observed 
for the TDA dynamic.\cite{BUEN06} At $T=0.6T_c$,  the velocity is nearly 
isotropic for weaker fields, while becoming Eden like for stronger fields. 
The exception is the case $H/J=2$, which at small angles presents 
a polynuclear-growth type, as well as significant differences between
the theoretical and simulated results. 
The behavior of the normal velocity at $T=0.6T_c$ 
(excluding the case $H/J=2$), is very similar to that observed for both 
the TDA and the OSD dynamics.\cite{BUEN06} 

\begin{figure}[ht]
\includegraphics[angle=0,width=.50\textwidth]{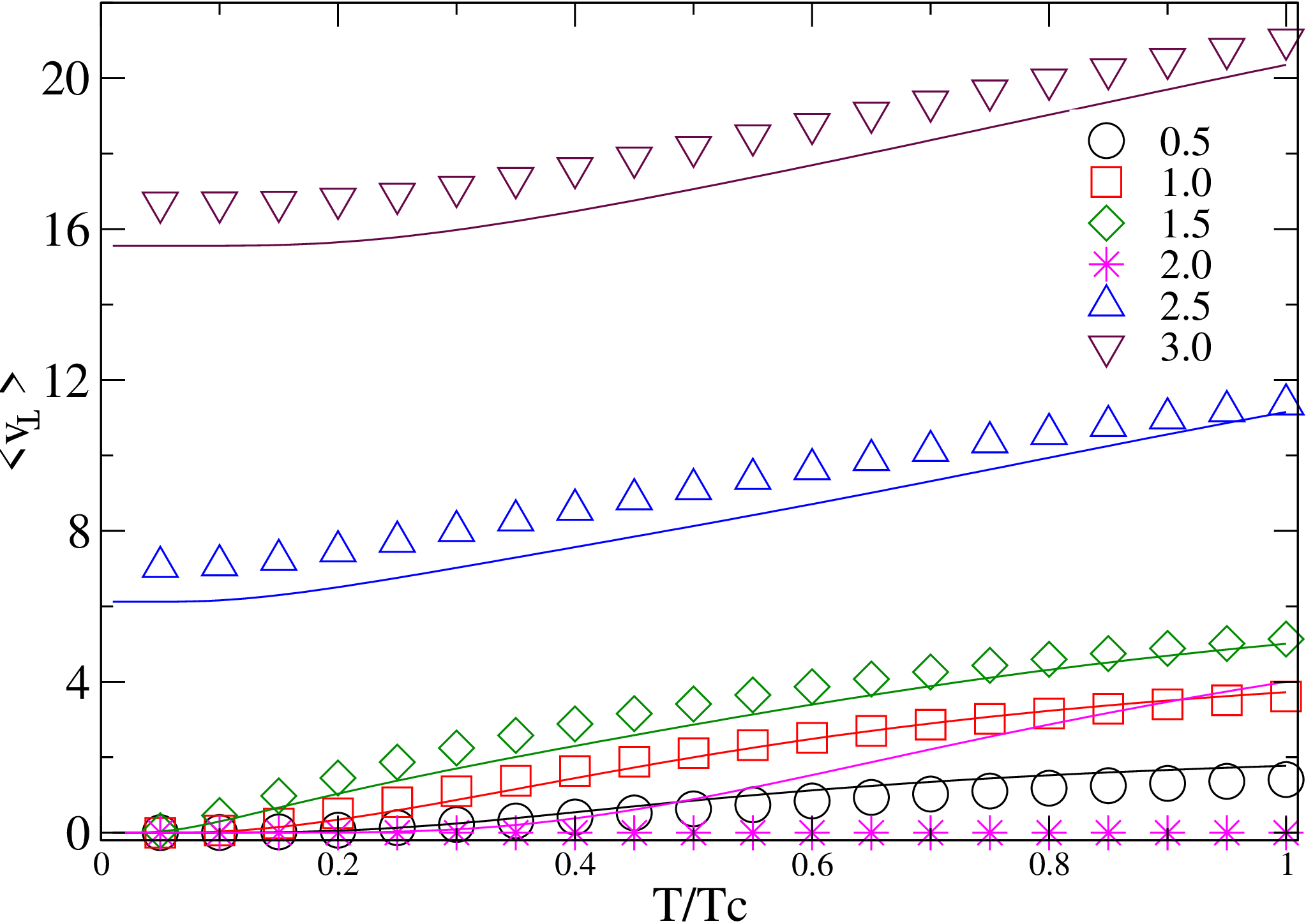}\\
\caption[]{
(Color online) The average stationary
normal interface velocity $\langle v_\perp \rangle$ vs $T$
for $\phi = 0$ and $H/J$ between $0.5$ and $3$. MC data are represented
by data points
and analytical results by solid curves. 
 Online, the colors of the curves and symbols match. The agreement between
 simulation and theory is quite good, except for $H/J=2$. 

}
\label{fig:vt}
\end{figure}

The temperature dependence of the normal interface velocity is shown in 
Fig.~\ref{fig:vt} for several values of $H/J$. 
The agreement between the simulations and the analytical results is
reasonable, 
except for $H/J=2$, where the simulated velocity remains zero for all 
temperatures, while the predicted velocity increases monotonically 
with the temperature. 
This discrepancy is also due to the metastable, static and microscopically 
flat, interface that forms at $H/J=2$.  
This figure also shows that as $T \rightarrow 0$, the system develops
a step discontinuity: 
the velocity is zero for $H/J \leq 2$ and increases with $H$ for stronger 
fields.
This discontinuity at $T=0$ is also observed with the TDA\cite{BUEN06} and 
with the standard Glauber dynamic (see Fig.~8 of Ref.~\onlinecite{RIKV02B}).

\subsection{Spin-class populations and skewness}
\label{sec:MCc}

\begin{figure}[ht] 
\begin{center}
\includegraphics[angle=0,width=.47\textwidth]{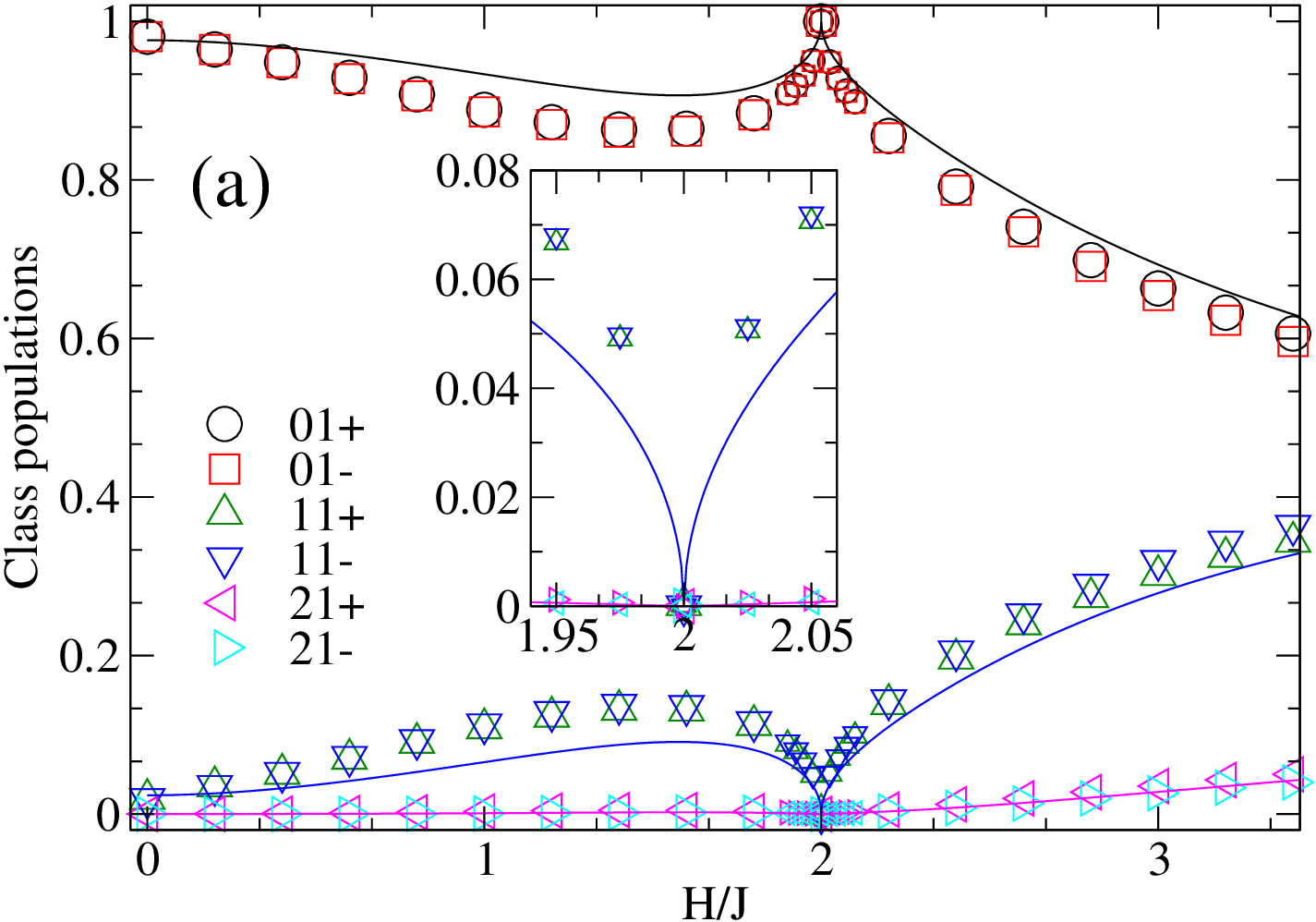}
\hspace{0.5truecm} 
\includegraphics[angle=0,width=.47\textwidth]{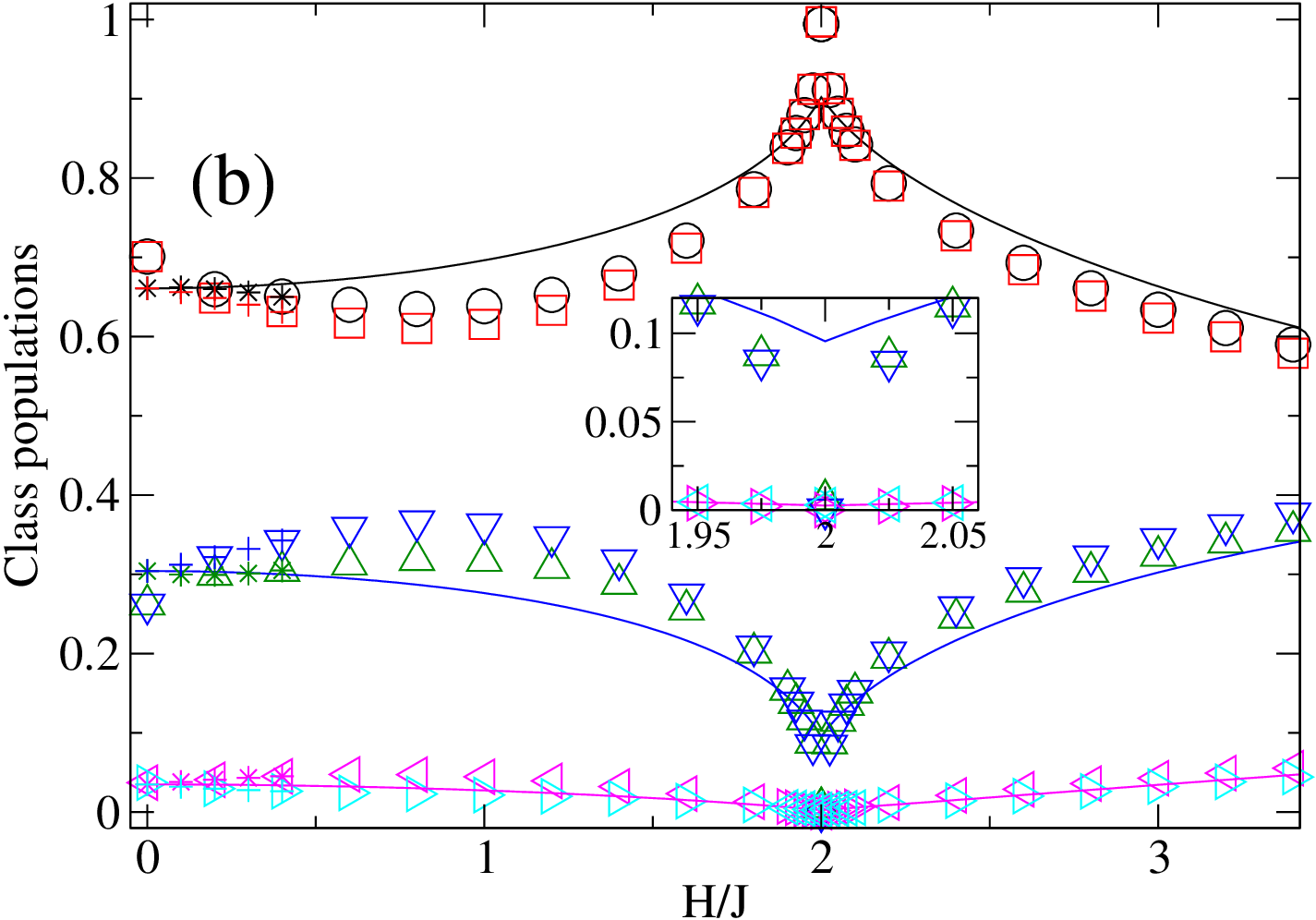}
\end{center}
\caption[]{(Color online)
Mean stationary class populations $\langle n(jks) \rangle$ vs $H/J$ 
for $\phi = 0$.
The simulation results are indicated by symbols, and the analytic 
approximations by solid curves. (a) $T=0.2T_c$. (b) $T=0.6T_c$. 
At $T=0.6T_c$ we also  
include some values (asterisks for $\langle n(jk+) \rangle$ and 
pluses for $\langle n(jk-) \rangle$ ) calculated with the interface 
created by the phonon-assisted dynamic, 
using the thermalized interface obtained by the Glauber dynamic 
as starting state. Note that in this case there is excellent 
agreement between theory and simulations at $H/J=0$.  
The other symbols have the same interpretations in (a) and (b), given by
the legend in (a).
The insets in both (a) and (b) show $\langle n(11s) \rangle$ and 
$\langle n(21s) \rangle$ near $H/J=2$. 

}
\label{fig:pop}
\end{figure}

To test the analytical assumption that different steps are statistically 
independent, we compare the analytical results for the mean class 
populations \cite{RIKV02B} with the simulated ones. The six mean class 
populations --- $\langle n(01s) \rangle$, 
$\langle n(11s) \rangle$, 
and $\langle n(21s) \rangle$ with $s = \pm 1$ --- 
for $\phi = 0$ at $T=0.2T_c$ and $T=0.6T_c$ are shown vs $H$ 
in Fig.~\ref{fig:pop}. At both temperatures, the 
analytical approximations follow the average of the populations 
for $s=+1$ and $s=-1$ qualitatively well. However, for small fields, $H/J<2$,
the simulations show a stronger dependence on $H$ than the mean-field
results. This  difference is more evident at the higher temperature, where
the simulations show that the population in front of the interface
($s = -1$) is quite different from the one behind it ($s=+1$). 
Well away from the special fields ($H/J=0$ and 2), the interfaces are a 
little rougher than the theory predicts (lower $01s$ and higher $11s$ 
populations). Near the special fields the interfaces appear to get caught 
in smoother (metastable) configurations. For $H=0$, the $01s$
populations are 
slightly higher that the theoretically predicted value, and the $11s$
populations are 
slightly lower. This is more evident at the higher temperature. Note
that when $H/J=0$ the interface should be
in equilibrium, and the theoretical value is exact.
However, this exact value is only reached if the interface is 
equilibrating properly. This can be seen in Fig.~\ref{fig:pop}(b) where we 
include the population averages, both calculated starting from a
microscopically flat interface and from a thermalized interface
generated with the standard Glauber dynamic. For $H/J=2$ the interface is much 
smoother than predicted. (For $T=0.6T_c$, the measured values for 
$\langle n(01\pm) \rangle$, $\langle n(11-) \rangle$, 
and $\langle n(11+)\rangle$ are approximately 
$0.9999$, $10^{-4}$, and $0$, compared with the respective predicted values of 
$0.9$, $0.1$ and $0.01$.)

\begin{figure}[ht] 
\includegraphics[angle=0,width=.50\textwidth]{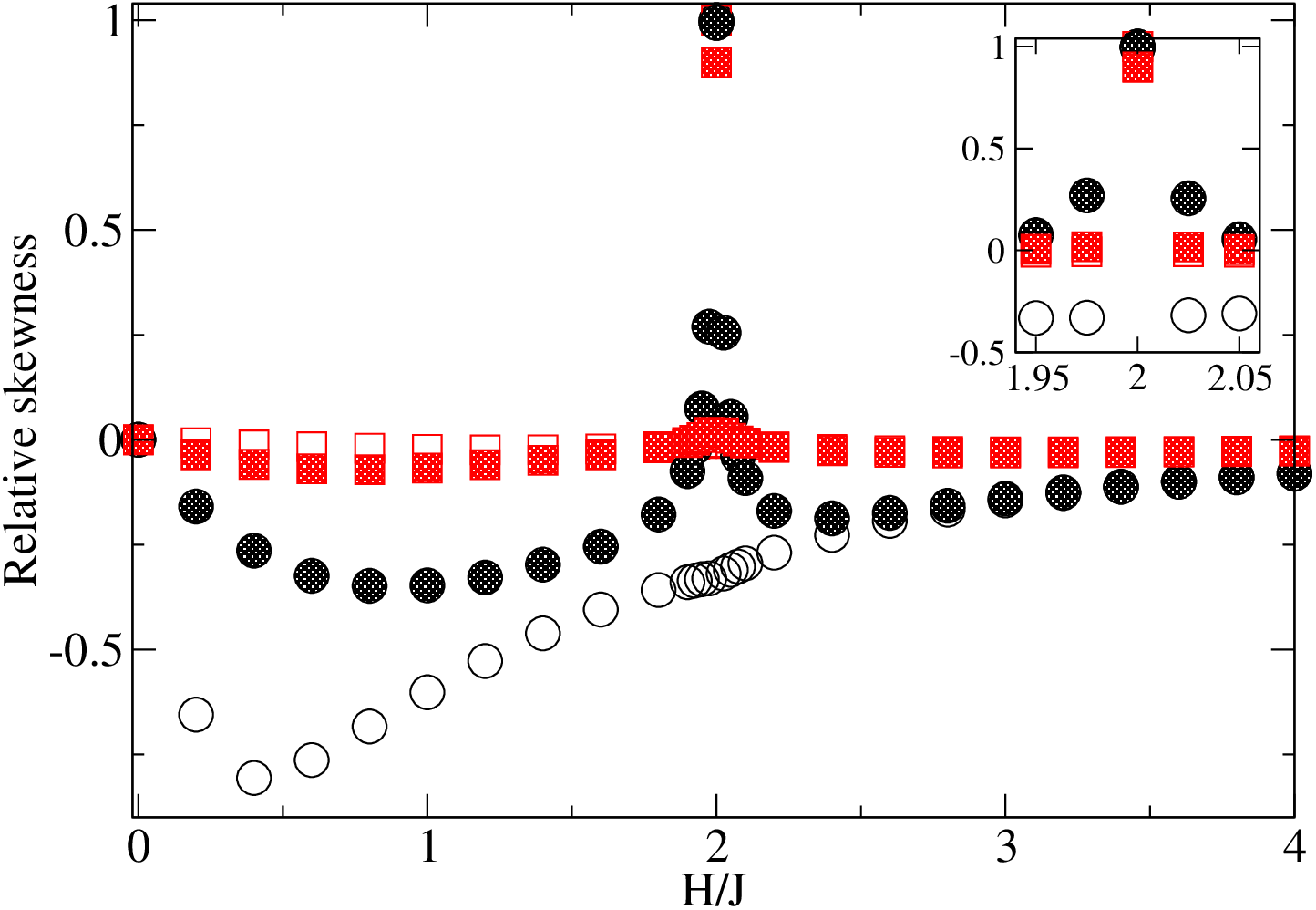}\\
\vspace{1truecm}
\caption[]{
(Color online) The two relative skewness parameters $\rho$ (circles, black)  and 
$\epsilon$ (squares, red online), defined in 
Eqs.~(\protect\ref{eq:rho}) and~(\protect\ref{eq:epsi}), respectively. 
The parameters are shown vs $H/J$ 
for $\phi = 0$, at $T=0.2T_c$ (empty symbols) and at $T=0.6T_c$ (filled
symbols). The inset shows a magnified view of the region around $H/J=2$. 
}
\label{fig:skew}
\end{figure}

The short-range correlations between neighboring steps are responsible for the skewness between the spin populations on the leading and trailing edges 
of the interface that appears in the simulation results. This phenomenon is commonly observed in driven interfaces. 
It occurs 
even when the {\it long-range\/} correlations vanish as they do
for interfaces in the 
KPZ dynamic universality class, to which the present model belongs for
all finite, nonzero values of $H/J \neq 2$. 
 Skewness 
has also been 
observed in several other SOS-type models, such as the body-centered SOS model 
studied by Neergaard and den~Nijs,\cite{NEER97} the model for step propagation 
on crystal surfaces with a kink-Ehrlich-Schwoebel barrier studied 
by Pierre-Louis et al.,\cite{PIER99} and a model for the local time 
horizon in parallel kinetic MC simulations studied by Korniss 
et al.\cite{KORN00C} 
No skewness was observed for the SOS model with the soft Glauber 
dynamic.\cite{RIKV02} However, some skewness was present in the
OSD model, \cite{BUEN06} indicating that a complete  lack of skewness is not a necessary characteristic of soft dynamics. Also, a small 
degree of skewness was observed for the Ising model (whose
interfaces include bubbles and overhangs) with the soft Glauber
dynamic (about two orders of magnitude smaller than the skewness observed for 
the hard Glauber dynamic).\cite{RIKV03} The TDA dynamic also presents 
considerable skewness. \cite{BUEN06}
The correlations associated with the skewness generally lead to a broadening 
of protrusions on the leading edge (``hilltops''), while 
those on the trailing edge (``valley bottoms'') are sharpened,\cite{NEER97} 
or the other way around.\cite{KORN00C} In terms of spin-class populations, 
the former corresponds to $\langle n(21-) \rangle > \langle n(21+) \rangle$ 
and $\langle n(11+) \rangle > \langle n(11-) \rangle$. The relative skewness 
can therefore be quantified by the two functions,\cite{NEER97} 
\begin{equation}
\rho = \frac{\langle n(21-) \rangle - \langle n(21+) \rangle}
{\langle n(21-) \rangle + \langle n(21+) \rangle}
\;,
\label{eq:rho}
\end{equation}
and\cite{RIKV02B}
\begin{equation}
\epsilon = \frac{\langle n(11+) \rangle - \langle n(11-) \rangle}
{\langle n(11+) \rangle + \langle n(11-) \rangle}
\;.
\label{eq:epsi}
\end{equation}
These two skewness parameters are shown together in Fig.~\ref{fig:skew}.
The temperature dependence of the skewness is stronger at the lower
temperature and smaller fields, and it is especially pronounced for
$\rho$, due to the low concentration of sites in the class $21+$ at low
temperatures and weak fields. $\epsilon$ is very small and almost
independent of $T$ and $H$, except at $H/J=2.0$, where both values are
near unity, consistent with the picture
of an interface with a very low density of $21-$ notches.

\section{Discussion and Conclusions}
\label{sec:DISC}

In this paper we have studied the nanostructure of an unrestricted 
SOS interface interacting with a two-dimensional phonon heat bath and driven far from equilibrium by an applied field. This work is
a continuation of previous studies aimed to explore the crucial role of
the stochastic dynamics selected to simulate physical 
systems.\cite{RIKV00B,RIKV02B,RIKV02,RIKV03,BUEN06,PARK04,BUEN04A} 
Important properties such as nanostructure and
mobility of driven interfaces have been shown to be strongly dependent
on the stochastic dynamics.\cite{RIKV00B,RIKV02B,RIKV02,RIKV03,BUEN06} 
 
The transition rates that give the evolution of the system are
calculated by coupling the spin Hamiltonian to a two-dimensional phonon
heat bath. The dynamic generated in this way belongs to the class known
as hard. \cite{MARR99} In condensed-matter systems, phonon-driven 
dynamics are dominant, and technologically they are becoming increasingly 
important, e.g., in connection with the behavior of quantum dots.\cite{MEUN06}

We studied the nanostructure and velocity of the SOS interface by kinetic MC simulations and by a non-linear
mean-field theory developed in previous papers.~\cite{RIKV00B,RIKV02B} 
 We calculated the interface velocity as a function
of the driving field, temperature, and angle of the interface relative to the lattice axes. We also studied the
local shape of the interface in terms of the spin-class populations, the average height of 
a step, and the probability density
for individual steps in the interface. 

In general we found good agreement between the theoretical calculations and the MC simulations. In particular we found the
strong dependence of the interface structure on the field characteristic 
of systems that evolve under hard dynamics, such as the Glauber or 
TDA dynamics.\cite{RIKV00B,RIKV02B, BUEN06} 

Our theoretical results are based on the mean-field assumption that 
individual steps of the
interface are statistically independent; short-range correlations are
neglected. However, our MC results show asymmetry between the spin populations
on the leading and trailing edges of the interface, 
which is an indication of the existence of such
short-range correlations. With increasing field, 
the interfaces undergo a gradual breakdown of up-down symmetry,
which has also been observed in 
other examples of driven interfaces.\cite{KORN00C,NEER97,PIER99,BUEN06} 
Aside from such, relatively minor, discrepancies between the
theoretical mean-field predictions and the simulation results, which 
show that there is room for improvement of
the mean-field model, the theory predicts
very accurately the qualitative behavior of the interfaces and yields a 
reasonable over-all approximation to their quantitative behavior. 
The important exceptions are the special field values, $H/J=0$ and 2, where
certain transitions allowed by the SOS constraint are forbidden by the
phonon-assisted dynamic for phonon baths of dimension greater than one. At $H/J=0$ this leads to a failure of the simulated
interface to reach thermal equilibrium, while at $H/J=2$ it leads to a
metastable, abnormally flat interface that is unable to propagate. 

It should, however, be noted that the phonon-assisted dynamic defined by
Eq.~(\ref{eq:wpb}) is based on a {\it weak, linear\/} coupling of the bosonic
bath to the spins. It is therefore possible that higher-order and/or
multiphonon 
corrections to the transition rates could restore the vanishing rates for 
energy-conserving transitions. Nevertheless, higher-order effects may not 
completely mask the slowing-down of the interface in the 
vicinity of the special field values observed here. We therefore expect 
that much of the characteristic field dependence will carry over to more 
sophisticated rate models. 
This expectation is supported by the results of recent experiments on
phonon-mediated spin dynamics in a quantum dot,\cite{MEUN06}
in which significant slowing-down was observed for nearly
energy-conserving transitions. 
Another interesting question is to what extent a relaxation of the SOS
constraint to consider a full Ising interface including overhangs and
bubbles (see Refs.~\onlinecite{RIKV00B} and~\onlinecite{RIKV03}) 
might open alternative
channels for full equilibration. Naturally, one can also think of more 
general dynamics, including multiple-spin-flip elementary transitions. 
These questions are left for future study. 

As in previous
studies,\cite{RIKV00B,RIKV02,RIKV02B,RIKV03,BUEN06,PARK04,BUEN04A}
our results indicate strong differences between interfaces moving under 
different stochastic dynamics, emphasizing the need for extreme care in 
selecting the appropriate dynamic for the physical system of interest. 
This general understanding and the specific results for the
phonon-assisted dynamic presented in this paper represent significant steps
in the direction of putting kinetic MC simulations on a solid physical
foundation.

\section*{Acknowledgments}
\label{sec:ACK}

G.~M.~B.\ appreciates the hospitality of the 
School of Computational Science at Florida State University. 
This research was supported in part 
by National Science Foundation Grant Nos.~DMR-0240078 and DMR-0444051,
 by Florida State 
University through the Center for Materials Research and Technology and
the School of Computational Science, by  the National High Magnetic Field Laboratory, and
by the Deanship of Research and Development of Universidad Sim{\'o}n Bol{\'{\i}}var.




%
\begin{table}[ht]
\caption[]{
The spin classes in the anisotropic square-lattice SOS model. 
The first column contains the class labels, $jks$. There are two other classes, $10s$ and $20s$,
that also have nonzero populations in the 
SOS model, but are not included because flipping a spin in any of them would produce an overhang 
or a bubble and is therefore forbidden.  
The second column contains the change in the total system energy
resulting from reversal of a spin from $s$ to $-s$, $\Delta E(jks)$.
The third column contains the mean spin-class populations for 
general tilt angle $\phi$, with $\cosh \gamma(\phi)$ from Eq.~(\ref{eq:chgam}). 
}


\begin{tabular}{| l | l |l|}
\hline
Class, $jks$
& $\Delta E(jks)$
& $ \langle n(jks) \rangle $

\\
 \hline\hline
 $01s$
 & $ 2sH + 4J_x $
& $\frac{1 - 2X \cosh \gamma (\phi) + X^2}{(1-X^2)^2}$

\\
 \hline
 $11s$
 & $ 2sH $
& $\frac{2X[(1+X^2) \cosh \gamma (\phi) - 2X]}{(1-X^2)^2}$

\\
 \hline
 $21s$
 & $ 2sH - 4J_x $
 & $\frac{X^2[1-2X\cosh\gamma(\phi)+X^2]}{(1-X^2)^2}$

\\
\hline
\end{tabular}
\label{table:class}
\end{table}

\vspace{2.0truecm}


\end{document}